	\documentclass[journal,twocolumn,twoside,a4paper,10pt
	]{IEEEtran}
	\usepackage[square,comma,compress,numbers]{natbib}
	\usepackage[english]{babel}  
		\let\ORGforeignlanguage\foreignlanguage
		\def\foreignlanguage#1{\lowercase{\ORGforeignlanguage{#1}}} 
	\usepackage[latin1]{inputenc}
	\usepackage[dvips%
	]{graphicx}
	\usepackage{amssymb}
	\usepackage{amsmath}
	\usepackage{epsfig}
	\usepackage{color} 
	\usepackage[raggedright,tight]{subfigure}
	\usepackage{booktabs}

		\newcommand{\maybecoloreqn}[1]{{#1}}

\newcommand{\cit}[1]{\,\cite{#1}}

\newcommand{\hh}{\ensuremath{\mathfrak{h}}}
\newcommand{\vr}{\vx}
\newcommand{\vx}{\ensuremath{\myvecx}}
\newcommand{\myvecx}{\myvec{r}}
\newcommand{\myvec}[1]{\vec{#1}}

\newcommand{\Real}{\mathfrak{Re}}
\newcommand{\Imag}{\mathfrak{Im}}

\newcommand{\HH}{\protect{\ensuremath{\text{H}}}}
\newcommand{\LL}{\protect{\ensuremath{\text{L}}}}

\begin{document}
%
\title{\bf{Post-Processing Enhancement of Reverberation-Noise Suppression in Dual-Frequency {SURF} Imaging}}
\author{Sven~Peter~{Näsholm},~\IEEEmembership{Member,~IEEE}, Rune~{Hansen}, and~Bjørn~A.~J.~{Angelsen},~\IEEEmembership{Senior~Member,~IEEE} %
%
%
\thanks{This work was supported by the Medicine and Health program of the Research Council of Norway.}
\thanks{The research was done while the authors were with the Department of Circulation and Imaging, Norwegian University of Science and Technology, Trondheim, Norway. Sven Peter Näsholm is now with the Department of Informatics, University of Oslo, Norway (e-mail: svenpn@ifi.uio.no).}
\thanks{Rune Hansen is also with SINTEF Health Research, Trondheim.}%
\thanks{Digital Object Identifier 10.1109/TUFFC.2011.1811}
}

\markboth{\sc E-print. IEEE Trans.\ Ultrason., Ferroelectr., Freq.\ Control, vol.~58, no.~2, February 2011}{Näsholm \textit{\MakeLowercase{et al.}}: {Post-Processing Enhancement of Reverberation-Noise Suppression in {SURF} Imaging}}

\maketitle
	\newlength{\figwidthC}
	\setlength{\figwidthC}{.46\columnwidth}
	\newlength{\figwidthCa}
	\setlength{\figwidthCa}{.49\columnwidth}
	\newlength{\figwidthCb}
	\setlength{\figwidthCb}{.49\columnwidth}
	\newlength{\figwidthCc}
	\setlength{\figwidthCc}{.49\columnwidth}
	\newlength{\figw}
	\setlength{\figw}{.45\columnwidth}
	\newlength{\figwidthCf}
	\setlength{\figwidthCf}{.45\columnwidth}

\newcommand{\linedescription}{\includegraphics{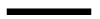}: HF pressure non-adjusted, \includegraphics{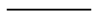}: HF difference with time-shift adjustment, \includegraphics{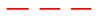}: HF difference with filter adjustment $\hh_{z_a}$}
\newcommand{\panedescription}{Left: HF difference with $\hh_{z_a}$ adjustment, middle: HF difference with $\tau_a$ adjustment, right: HF pressure non-adjusted.}
\begin{abstract}
A post-processing adjustment technique which aims for enhancement of dual-frequency SURF (Second order UltRasound Field) reverberation-noise suppression imaging in medical ultrasound is analyzed. Two variant methods are investigated through numerical simulations. They both solely involve post-processing of the propagated high-frequency (HF) imaging wave fields, which in real-time imaging corresponds to post-processing of the beamformed receive radio-frequency signals. %
Hence the transmit pulse complexes are the same as for the previously published SURF reverberation-suppression imaging method. %
The adjustment technique is tested on simulated data from propagation of SURF pulse complexes consisting of a 3.5\,MHz HF imaging pulse added to a 0.5\,MHz low-frequency sound-speed manipulation pulse. Imaging transmit beams are constructed with and without adjustment. %
The post-processing involves filtering, \emph{e.g.} by a time-shift, in order to equalize the two SURF HF pulses at a chosen depth. This depth is typically chosen to coincide with the depth where the first scattering or reflection occurs for the reverberation noise one intends to suppress. %
The beams realized with post-processing show energy decrease at the chosen depth, especially for shallow depths where in a medical imaging situation often a body-wall is located. This indicates that the post-processing may further enhance the reverberation-suppression abilities of SURF imaging. %
Moreover, it is shown that the methods might be utilized to reduce the accumulated near-field energy of the SURF transmit-beam relative to its imaging region energy. The adjustments presented may therefore potentially be utilized to attain a slightly better general suppression of multiple scattering and multiple reflection noise compared to for non-adjusted SURF reverberation-suppression imaging.
\\

\noindent\verb#The peer-reviewed version of this paper is published #\\
\verb#in IEEE Transactions on Ultrasonics, Ferroelectrics #\\
\verb#and Frequency Control, vol. 59, no. 2, pp. 338-348,#\\
\verb#February 2011. DOI: 10.1109/TUFFC.2011.1811 #\\
\verb#The final version is available online at #\\
\verb#http://dx.doi.org/10.1109/TUFFC.2011.1811 The current# \\
\verb#e-print is typeset by the authors and differs in e.g.#\\
\verb#pagination and typographic detail.# 

\end{abstract}
\IEEEpeerreviewmaketitle

\section{Introduction}
\IEEEPARstart{T}{he} SURF imaging synthetic transmit-beam for rever\-beration suppression, below referred to as the SURF beam, is generated from the difference between two subsequent ultrasound imaging pulses transmit in the same direction\cit{nasholm2009}. Each of the transmissions consists of a dual-frequency pulse complex with a conventional high-frequency (HF) imaging pulse added to a low-frequency (LF) tissue manipulation pulse. The high frequency $f_\text H$ may typically be around 10 times the low frequency $f_\text L$. The LF polarity is switched for the second transmission so that the LF pressure experienced by two HF pulses becomes of opposite polarity. Therefore, due to the nonlinear effect of pressure-dependent compressibility of tissue, the LF-pulse dependent effective speed of sound experienced during propagation of these two HF pulses differs. %
As the covered distance increases, a difference is thus accumulated between the propagation time required for the HF pulses of the two transmissions. The difference obtained when subtracting the two propagated HF pulse components hence grows with covered distance until they possibly are separated by a delay corresponding to a $180^\circ$ phase-shift. %
This effect may be utilized during image reconstruction in order to suppress multiple scattering and multiple reflection artifacts (in the rest of this paper considered as and denoted as reverberation artifacts), as described and analyzed in Ref.~\citenum{nasholm2009} where also combined SURF and pulse-inversion image reconstruction is introduced. In the citet paper, SURF transmit beams are simulated for two different pulse and transducer setups and compared to conventional fundamental and pulse-inversion second-harmonic transmit-beams. In this paper, the theory behind SURF reverberation suppression is further described in Section \ref{section:theory_methodsC}. For a recent review of tissue nonlinearity, second-harmonic tissue imaging, and related research, see Ref.~\citenum{duck2010}. %
\IEEEpubidadjcol 

The reverberation suppression ability of SURF imaging is illustrated by its synthetic transmit beam being reduced near the transducer, as is also the second-harmonic transmit-beam in tissue harmonic imaging\cit{thomas:thi_why_does_it_work}. %

Dual-frequency pulses have also been used for ultrasound contrast-agent microbubble detection\cit{deng2000, hansen:phd, bouakaz:radial_modulation, masoy:SURF_in_vivo, angelsen_hansen:ieee_proc07, hansen2009, emmer2009}. Within this context, SURF imaging is sometimes denoted radial modulation imaging. %
The contrast agent detection has been shown \emph{in vivo}\cit{masoy:SURF_in_vivo}. %
In Ref.~\citenum{hansen2010}, the versatility of SURF imaging is illustrated, and fundamental concepts in relation to the use of dual-frequency pulse complexes in ultrasound imaging are investigated both regarding nonlinear propagation and nonlinear scattering. A variant of the time-delay post-processing adjustment analyzed in the present article is also considered. %
Furthermore, dual-frequency pulse insonification has been employed for imaging the nonlinearity parameter of different media\cit{ichida1984, sato1985, cain1986, cain1989, fukukita1996, pasovic2007}.
%
%
%


The purpose of the adjustment methods analyzed in the present paper is to further improve SURF reverberation artifact suppression, especially if the dominating rever\-beration noise {is} due to a strong first scattering/reflection taking place around a depth $z_a$ in the near-field. Such suppression would be helpful in clinical applications \emph{e.g.} where some weakly scattering anatomic feature is to be studied within a hypoechoic region. Then strong rever\-beration noise artifacts risk to camouflage the features intended to be visualized. %

The transmitted and propagated pulse complexes utilized are equal to those in non-adjusted SURF imaging. As described in Section~\ref{section:theory_methodsC}, a transmit synthetic SURF HF wave field is generated as a difference, with one of the propagated HF wave fields being adjusted, before the adjusted SURF difference is calculated. %
Subsequently a transmit beam is generated in a conventional way from the the root-mean-square (RMS) or the temporal maximum of the adjusted difference field. %

The rest of this article is organized as follows: %
First the SURF reverberation-suppression imaging method is reviewed. Then comes a description of three different kinds of time-shifts of importance for conceptual under\-standing of the here introduced {post-processing} method{s}. %
Two post-processing signal adjustments are depicted: adjustment by a time-shift and adjustment by a general filter. %
The introduction of th{e} general filter, which in addition to time-shifting also adjusts the pulse form, is motivated by the possibility that the two HF wave forms may be distorted if the experienced LF pressure varies along the HF pulse at a given spatial location. This variation is of opposite sign for switched LF polarity. %

Subsequently, a SURF beam gain-factor associated to the time-shift adjustment is explained. %
Then is defined a specific beam quality ratio $Q_{z_a}$ utilized to score transmit beams regarding suppression of reverberation noise stemming from a first scattering/reflection emerging at a depth $z_a$. %

The findings in the Results section are based on post-processing adjustment of data generated by nonlinear propagation simulation of two transmit SURF pulse complexes of opposite LF polarity. %
The transducer and pulse complex setup is described, then the effect on the spatial development of the gain factor due to {the} post-processing time-shift is shown. %
Then the beam-quality ratio of the calculated beams is provided. 
Finally, SURF beams are constructed and compared for non-adjusted data as well as data adjusted by time-shifts and by general filters.

\section{Theory and Methods}\label{section:theory_methodsC}
\subsection{SURF reverberation-suppression imaging outline\label{sec:above}}
This subsection describes SURF reverberation-suppression imaging without post-processing adjustment. %
Each {SURF} transmit pulse is a dual-frequency complex composed of a high-frequency (HF) imaging pulse added to a low-frequency (LF) tissue manipulation pulse. Reverberation noise suppression is achieved by taking advantage of an accumulated propagation time-delay of the HF pulse, introduced by the LF pulse. In the medical ultrasound regime, the nonlinear LF pressure-dependent speed of sound experienced by the HF pulse is\cit{nasholm2009, hansen2010}
\begin{align}
	c(\vr) = c_0(1+\beta_n\kappa\,p_\text{L}(\vr)),
	\label{eq:effective_c}
\end{align}
where $c_0$ is the speed of sound with no LF manipulation present, $\beta_n$ is the material nonlinearity parameter, $\kappa$ the compressibility and $p_\text{L}(\vr)$ is the LF pressure at the spatial location $\vr$. %
 The effective speed of sound when propagating in a positive LF pressure, and therefore compressed medium, thus becomes higher than in case of the medium being expanded due to a negative LF pressure. %

For SURF reverberation suppression image reconstruction, one transmits two SURF pulses for each transmit focal zone and direction. The HF pulse is equal for both transmissions, while the polarity of the LF pulse is switched for the second pulse. %
These HF pulses thus propagate at different effective speed of sound due to the opposite polarity of the associated LF pulses. Consequently a relative propagation time-shift $\tau(\vx)$ is generated between them, which is small at shallow depths, but is accumulated during forward propagation within the compressed/expanded medium. The synthetic SURF transmit beam is constructed from the difference between the {HF} wavefields. For a given $\vx$, its maximum amplitude is for the accumulated time-shift between the two HF pulses $\tau(\vx)=1/(2f_\HH)$. %

Due to the amplitude reduction of all reflected or scattered parts of a wave, an imaging pulse experiences negligible sound speed manipulation after being scattered, since the scattered or reflected LF wave it propagates in conjunction with is significantly weakened. Therefore the multiply scattered or reflected HF wave accumulates a smaller relative time-shift than the direct forward-propagating part. The reverberated parts of the receive HF pulses thus give negligible contribution to the receive SURF difference signal if the first scattering/reflection takes place near the transducer where the accumulated time-shift $\tau(\vx)$ during forward propagation is small\cit{nasholm2009}. The cited paper describes three different classes of reverberation noise, and the simulated beams are used to compare the reverberation suppression abilities of the respective imaging methods. This is done by comparison of an introduced quality parameter $Q$ (also stated here in Section \ref{sec:discussion}), which measures the ratio between the beam energy ratio within the imaging depth region and the beam energy closer to the transducer, thus quantifying how well Class\,I and II reverberation noise is suppressed. %
\subsection{Time-shift terminology}
There are three time-shifts of importance for  for conceptual understanding of the presented post-processing adjustment method:
{(}i) $\tau_0$ between the center of the HF and LF pulses in a transmit SURF pulse complex\label{enum:1}, see Ref.~\cit{nasholm2009,masoy:SURF_in_vivo} for further elaboration, %
{(}ii) $\tau(\vx)$ accumulated as the HF pulse propagates. It is due to the nonlinear sound speed change \eqref{eq:effective_c} induced by the LF part of the SURF pulse complex\label{enum:2}, as explained above in Section~\ref{sec:above}. %
{(}iii) the post-processing reverberation-suppression adjustment time-shift $\tau_a$, applied to one of the propagated HF fields before forming the adjusted SURF difference field, as described below in Section~\ref{subsec:tau_a}.


%
\subsection{SURF transmit-beam adjustment by a post-processing time-shift $\tau_a$\label{subsec:tau_a}}
In the model provided below, $s_+(\vx, t)$ represents the HF part of a SURF wave field propagated with positive LF manipulation polarity, while $s_-(\vx,t)$ represents the corresponding field in case of negative LF polarity. In this model, these HF wavefield components differ solely by a spatially varying accumulated time-shift $\tau(\vx)$ {induced by the LF sound speed manipulation}. In addition, a constant post-processing time-shift $\tau_a$ is applied to $s_-(\vx,t)$ to generate the field denoted $\hat s_-(\vx,t)$. The HF field after propagation without LF manipulation is denoted $s(\vx,t)$. Then $s_+(\vx, t) = s(\vx, t+\tau(\vx)/2)$ and $s_-(\vx, t)= s(\vx, t-\tau(\vx)/2-\tau_a)$. %
The time-shift adjusted SURF HF difference field then becomes:
\begin{align}
	s_\Delta(\vx,t) =&s_+(\vx,t)-\hat s_-(\vx,t)\label{eq:SURF_signal_pishift}\nonumber\\
	=&s\big(\vx, t+{{\tau(\vx)}/{2}}\big) - s\big(\vx, t-{{\tau(\vx)}/{2}}-\tau_a\big).
\end{align} %
For a non-adjusted field on the $s(\vx, t) = \Real\big\{\tilde s(\vx, t)e^{i\omega_0t}\big\}$ form, where $\tilde s(\vx, t)$ is a complex pulse envelope and $\omega_0$ the center angular frequency, the difference hence becomes 
\begin{align}
        s_\Delta(\vx,t) =\ & 
		\Real\big\{%
		+ \tilde s\big(\vx, t+{{\tau(\vx)}/{2}}\big)\,e^{i\omega_0(t+\tau(\vx)/2)}
 		\nonumber\\ 
	 	&- \tilde s\big(\vx, t-{{\tau(\vx)}/{2}} -\tau_a\big)\,e^{i\omega_0(t-\tau(\vx)/2-\tau_a)}%
 	\big\}.
\end{align}
Having the envelope $\tilde x$ unchanged during a time-shift, which is a narrow-band approximation, yields:
\begin{align}
        s_\Delta =& \Real\big\{\tilde s(\vx, t)e^{i \omega_0 t}\big[e^{i\omega_0\tau(\vx)/2}-e^{-i\omega_0(\tau(\vx)/2 +\tau_a)}\big]\big\}\nonumber\\
	=& \underbrace{2\sin[\omega_0(\tau(\vx)+\tau_a)/2]}_{\displaystyle \triangleq G(\omega_0,\tau(\vx)+\tau_a)} \Imag\left\{\tilde s(t)e^{i\omega_0t}\right\},
\label{eq:gain_factor_def}
\end{align} 
where $G$ may be interpreted as a spatially varying gain factor regulating the amplitude of $s_\Delta(\vx, t)$, relative to the non-shifted field $s(\vx, t)$. The temporal maximum of $s_\Delta$ may thus be estimated from knowledge of $\tau(\vx)$, $\tau_a$, and the temporal maximum of $s(\vx,t)$ without LF manipulation:
\begin{align}
	\max_t\Big| s_\Delta \Big| &= \max_t\Big| G(\omega_0,\tau(\vx)+\tau_a)\Imag\left\{ \tilde s(\vx, t)e^{i\omega_0 t} \right\}  \Big|\nonumber\\
	&=  \left|G(\omega_0, \tau(\vx)+\tau_a)\right|  \max_t\Big|\Imag\left\{ \tilde s(\vx, t)e^{i\omega_0 t} \right\}  \Big|,\nonumber
\end{align}
or equivalently for the RMS. The first maximum and minimum of $\left|G(\omega_0,\tau(\vx)+\tau_a)\right|$ are thus $\maybecoloreqn{\{}2,0\maybecoloreqn{\}}$ for $\omega_0(\tau(\vx)+\tau_a)  =\maybecoloreqn{\{}\pi, 0\maybecoloreqn{\}}$. %

With no time-shift adjustment applied, $G=0$ for $z=0$ (at the transducer), while application of $\tau_a$ adjusts the depth of maximum suppression. The time-shift adjustment $\tau_a$ giving maximum suppression of $s_\Delta(\vx,t)$ at the on-axis depth $z_a$ may be predicted from \eqref{eq:gain_factor_def} for $G(\omega_0,\tau(z_a)+\tau_a)) = 0$, which equally gives $s_+(z_a,t) = s_-(z_a,t)$. %


\subsection{SURF transmit-beam adjustment by a general post-processing filter}
The nonlinear LF manipulation might generate HF waveform change differing between $s_+(\vr,t)$ and $s_-(\vr,t)$ in a manner that cannot fully be described by a time-shift, \emph{e.g.} {due to} pulse compression or expansion. Then a time-shift adjustment $\tau_a(z_a)$ does not exist to make $s_+(\vr,t)=s_-(\vr,t)$ at $z=z_a$. However{, instead of a pure time-shift} a more general filter $\hh_{z_a}$ may be applied to $s(\vx,t)$ to make $s_+(\vx,t)=\hh_{z_a} \big\{s_-(\vx,t)\big\} (\vx, t)$ at the reference depth $z_a$. Then the SURF difference field becomes
\begin{align}
	s_\Delta(\vx,t) = s_+(\vx, t) - \hh_{z_a} \big\{s_-(\vx,t)\big\} (\vx, t).
	\label{eq:SURF_signal_filter}
\end{align}
In this work, $\hh_{z_a}$ is chosen as the Wiener filter making $s_+(\vr,t)=s_-(\vr,t)$ on-axis at the reference depth $z_a$. This filter is then applied to the whole $s_-(\vx,t)$ field. 

\subsection{Specific beam quality ratio}
For reverberation suppression in practical imaging, it is important that the sensitivity to multiple scattered and multiple reflected waves is low compared to the sensitivity to direct back-scattered waves from the imaging depth. Thus reduction of the transmit beam at the depth $z_a$ by a factor $C$ is of no use if the beam is equally suppressed by $C$ within the imaging region. %
To estimate the reverberation suppression characteristics of SURF transmit beams in case of a first scattering/reflection at the depth $z_a$, a specific beam quality ratio $Q_{z_a}$ is here introduced as the transmit beam energy within the imaging region divided by the summed beam energy at the depth $z_a$:
\begin{align}
	Q_{z_a}\triangleq {\displaystyle \sum_{z=z_n}^{z_f} \sum_{r=0}^{\infty}\sum_{\theta=0}^{2\pi} E(\vx)}  \left/\   {\displaystyle \sum_{r=0}^{\infty}\sum_{\theta=0}^{2\pi} E(\vx)}\big|_{z=z_a}\right.,
	\label{eq:quality_measureC}
\end{align}
where $E(\vx)$ is the beam energy at the spatial coordinate $\vx$ described by the cylindrical coordinates $(z,r,\theta)$. The imaging depth region is within $z\in[z_n,z_f]$ and is equal to the focal region given the active transmit aperture and imaging frequency. %

\section{Results}
\subsection{Simulation setup}
\begin{figure}[!tb]
\centering
  \includegraphics[width=\figwidthCf]{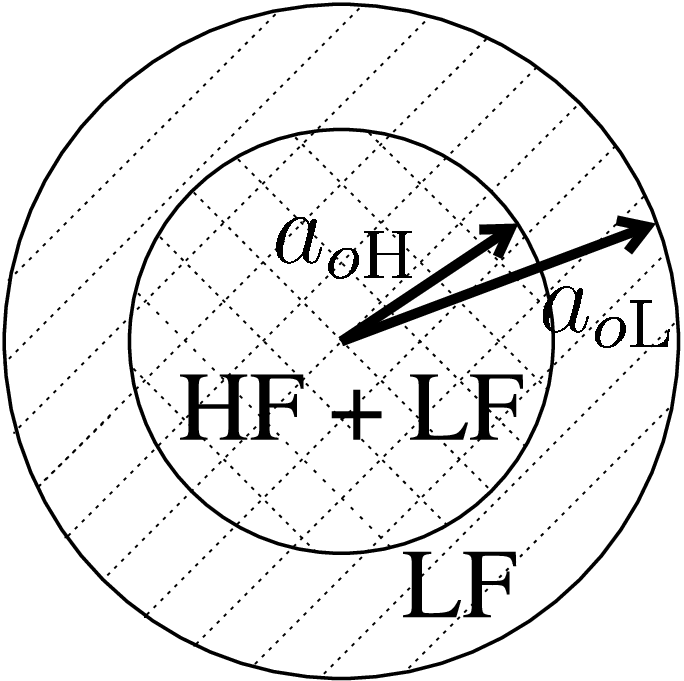} %
  \caption{Transmit aperture. Central region: HF and LF, outer region: only LF.}
  \label{fig:HFLF_versionsC}
\end{figure}

The raw data utilized to generate the results presented below is the same computer simulated fields of HF pulses, propagated in conjunction with LF manipulation of positive and negative polarity, as generated for the 3.5\,MHz setup in Ref.\:\citenum{nasholm2009}.  These HF wave fields are used as $s_+(\vx,t)$ and (non-adjusted) $s_-(\vx,t)$. The two adjustment variants (the pure time-delay filter and the delay and pulse-form filter) are applied to $s_-(\vr,t)$ to generate $\hat s_-(\vr,t)$. %
The modeled transmit aperture is axisymmetric consisting of a central region with both HF and LF transmission and an outer region where only LF is transmitted. Fig.~\ref{fig:HFLF_versionsC} shows the geometry of such HF and LF transmission regions. %
\begin{table}[!tbh]
	\caption{\label{tab:HF_aperturesC}Transmit pulse parameters for SURF beam generation.}
\begin{center}
\begin{tabular}{c c c c c c c c c c c c c c c c c c c}
	\toprule
$f_{\HH}$ & $\beta_\text{H,6dB}$ & $f_\text{L}$ & $\beta_\text{L,6dB}$ & $p_{0\HH}$ & $p_{0\LL}$ & $a_{o\HH}$\\ 
{\tiny [MHz]} & \tiny{$\%$} & {\tiny [MHz]} & \tiny{$\%$} & {\tiny [MPa]} & {\tiny [mm]}& {\tiny [mm]}\\
\midrule
$3.5$& $50$ & $0.5$ & $25$ & $3.5$ & $0.85$ & $7.1$ \\ 
\midrule
\end{tabular}
\begin{tabular}{c c c c c c c}
$a_{o\LL}$ & $F_\HH$ & $F_\LL$ & $\tau_0$ & $z_n$ & $z_f$\\
{\tiny [MPa]} & {\tiny  [mm]} & {\tiny [mm]} & {\tiny  [$\mu$s]} & {\tiny [mm]} & {\tiny [$\mu$s]}\\
\midrule
$10$ & $82$ & $82$ & $-0.2$ & $60$ & $130$\\
\bottomrule
\end{tabular}
\end{center}
$f_{\HH0}$ is the transmit HF frequency, $\beta$ the fractional frequency bandwidth, $p_0$ transmit surface pressure, $a_o$ the outer aperture radius, $F$ the focal depth, and $f_{\HH}$ the imaging frequency. $\HH$ and $\LL$ refer to high and low frequency. $\tau_0$ is the delay between HF and LF transmission. The imaging region is $z\in[z_n, z_f]$.
\end{table} %
The HF and LF frequencies, bandwidths, excitation pressures, and aperture radii are described in Tab.~\ref{tab:HF_aperturesC}, where also the {chosen} delay $\tau_0$ between transmission of the HF and LF pulses are given. The apertures are modeled without kerf. %
%
 
\subsection{Estimated time-shifts and gain factors}
\begin{figure*}[!tb]
	\centering
	\subfigure[$\tau_a=0$. Horizontally: depth, vertically: lateral coordinate.]{\makebox[\figwidthC][l]{\includegraphics[width=\figwidthC]{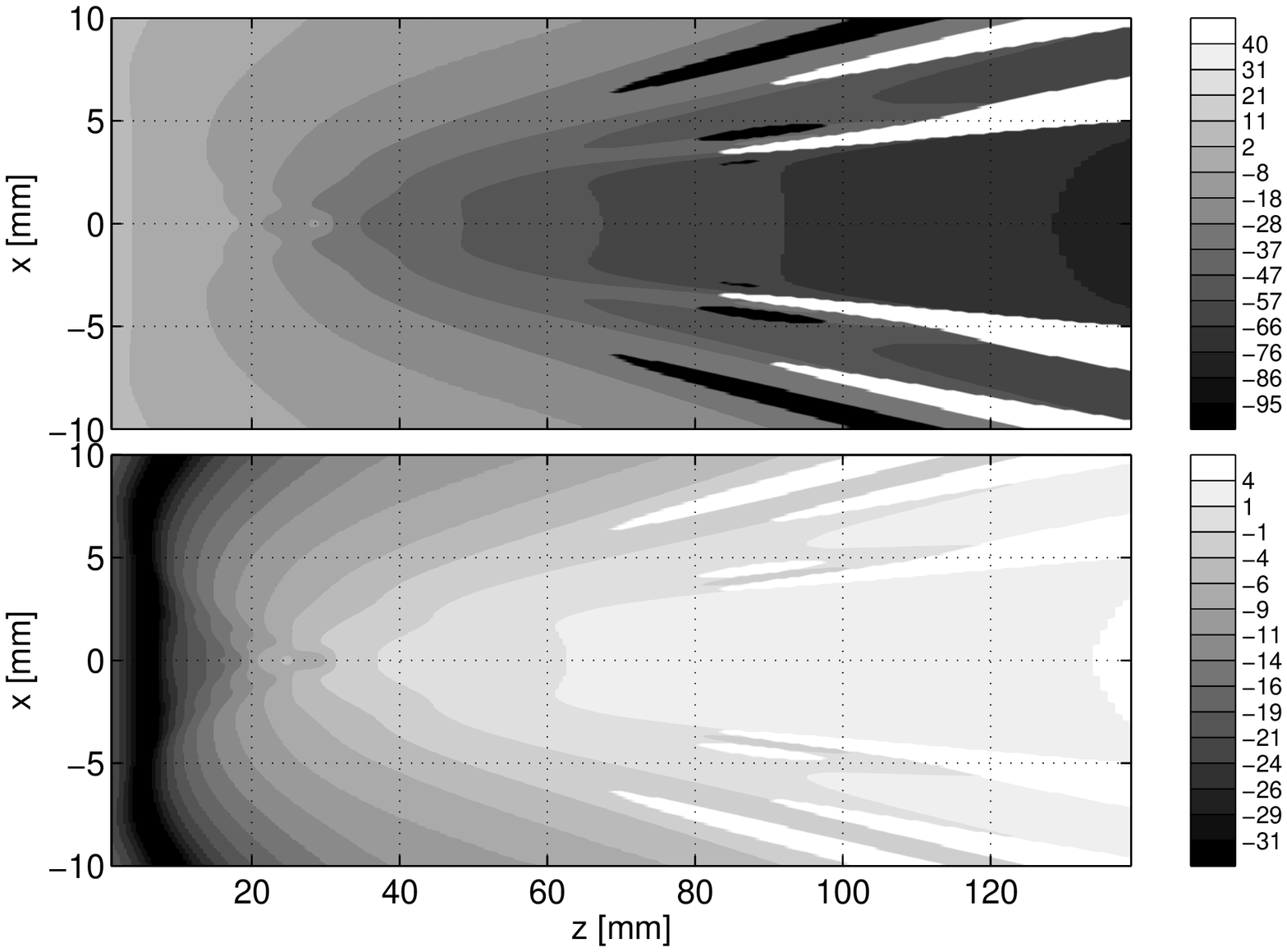}}}
	\subfigure[$\tau_a=7.1$\,ns. Axes as in (a).]{\makebox[\figwidthC][r]{\includegraphics[width=\figwidthC]{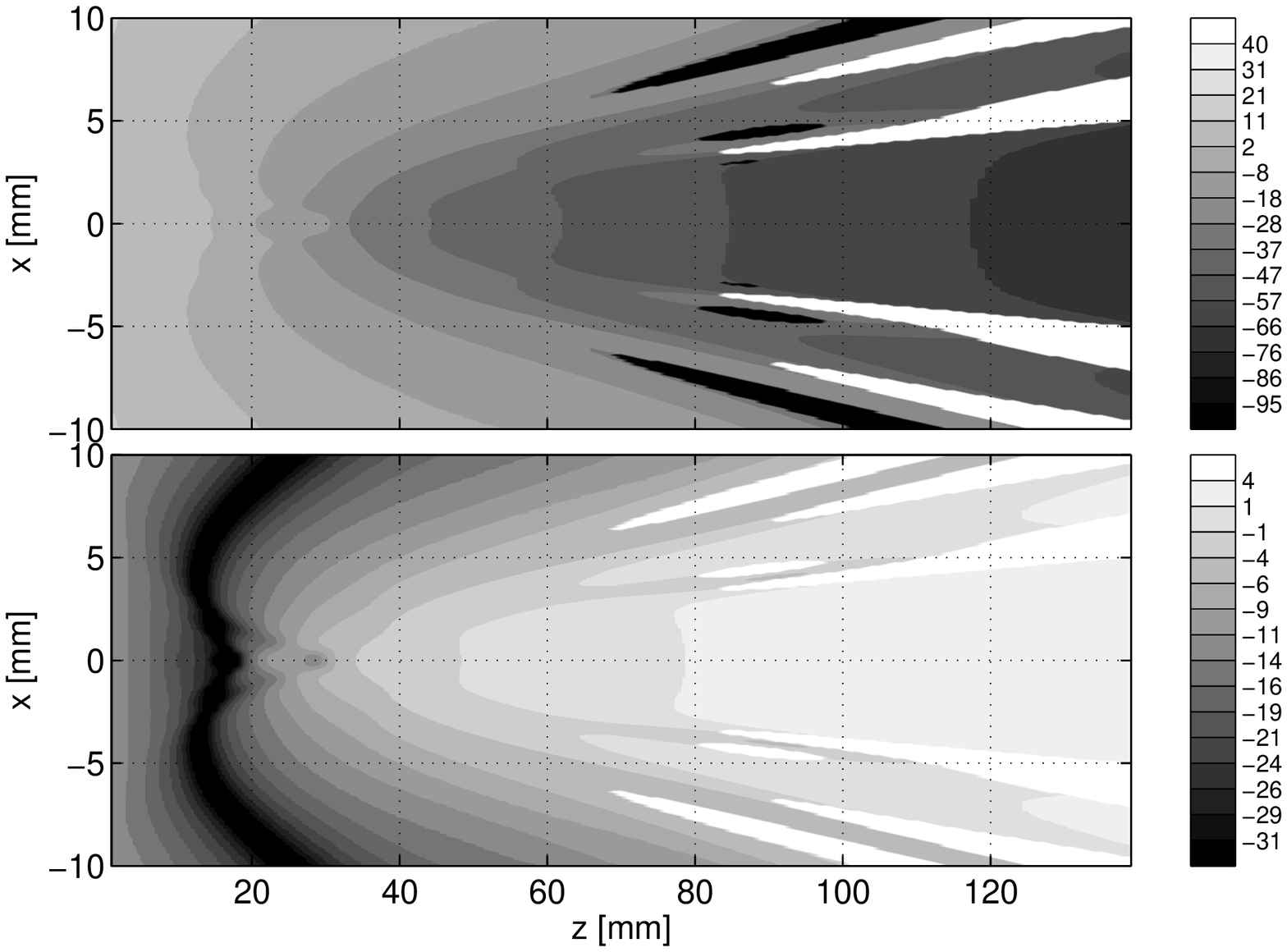}}}\\%
	\subfigure[$\tau_a=54$\,ns. Axes as in (a).]{\makebox[\figwidthC][r]{\includegraphics[width=\figwidthC]{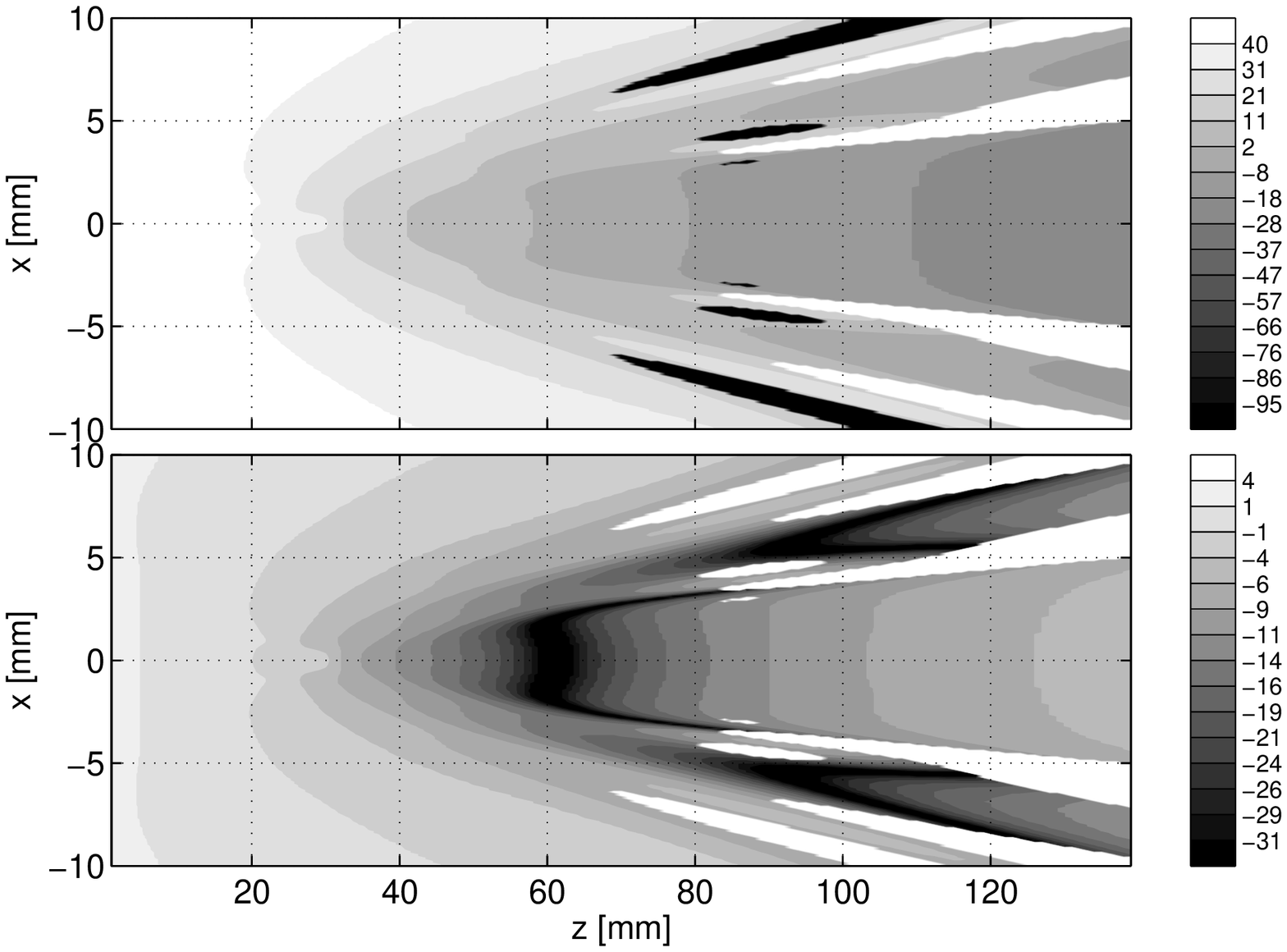}}}%
	\subfigure[On-axis $\tau(z)$ and $G(z)$.][On-axis $\tau(z)+\tau_a$ and corresponding $G(z)$. \includegraphics{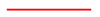}: $\tau_a=-21$\,ns, \includegraphics{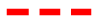}: $\tau_a=0$\,ns, \includegraphics{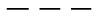}: $\tau_a=7.1$\,ns, \includegraphics{thick_line}: $\tau_a=54$\,ns.]{\makebox[\figwidthCa][c]{\includegraphics[width=\figwidthC]{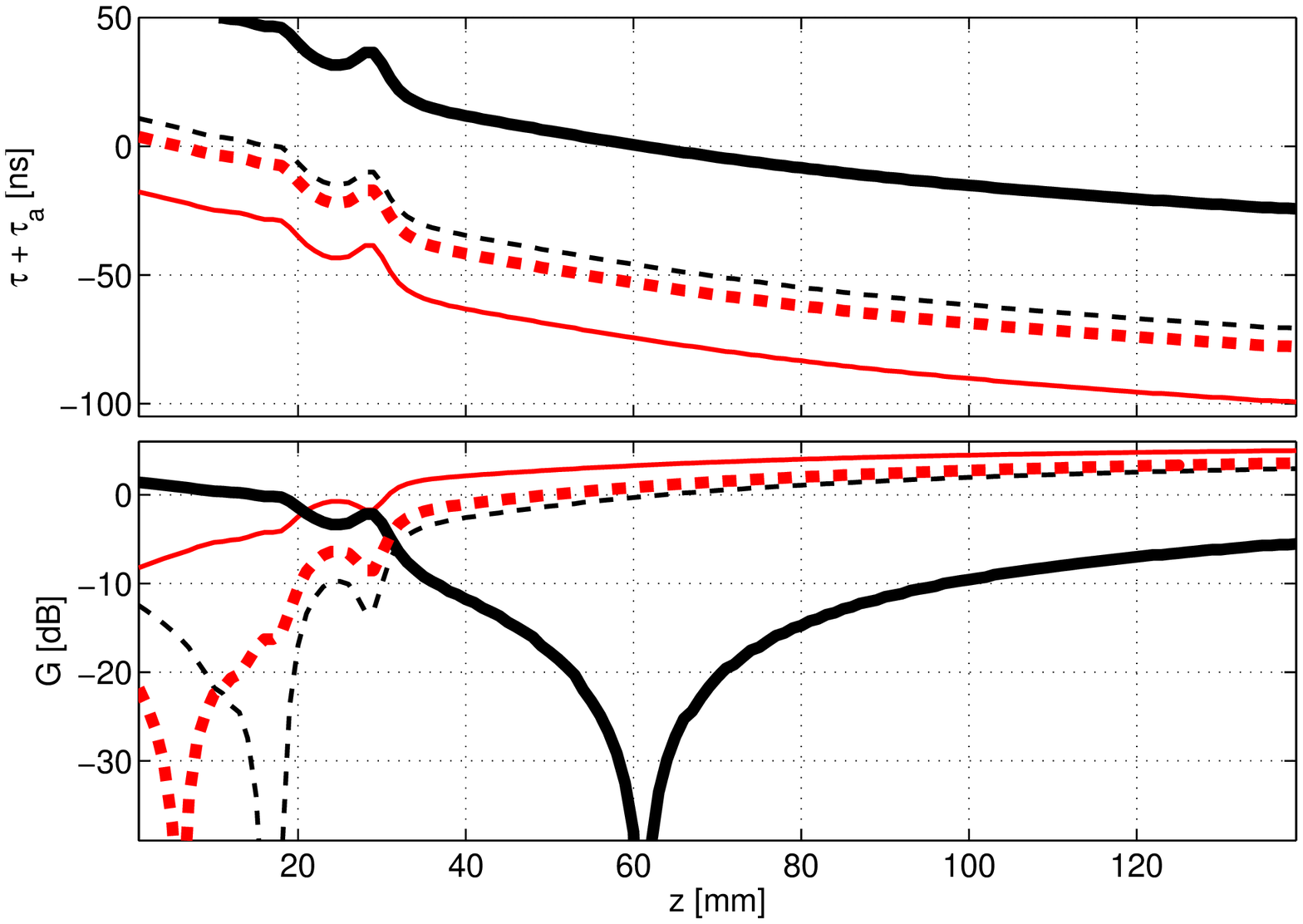}}}
	\caption{Top sub-panes: Estimated time-shift $\tau(z)+\tau_a$ [ns] between $s_+(\vx,t)$ and $\hat s_-(\vx,t)$. Bottom sub-panes: corresponding gain factor estimates $G(z)$ [dB].}
	\label{fig:gain_factors}
\end{figure*}

Fig.~\ref{fig:gain_factors} displays the estimated time-shift $\tau(\vx)$ between the simulated HF fields $s_+(\vx,t)$ and $s_-(\vx,t)$, as well as the corresponding gain factor estimates, calculated from the definition in Eq.\:\ref{eq:gain_factor_def}. %
The time-shift adjustment by $\tau_a$ corresponds to a vertical movement of the $\tau(z)$ curve, which influences the gain factor magnitude, especially within the near-field. To illustrate this, Fig\ \ref{fig:gain_factors} in addition shows $\tau(\vx)+\tau_a$ and $G(\omega_0,\tau(\vx)+\tau_a)$ for $\tau_a = \{-21, 7.1, 54\}$\,ns, corresponding to $\{-27^\circ, 9^\circ, 69^\circ\}$ phase-shifts at $3.5$\,MHz. %

\begin{figure}[!tb]
\centering
	\includegraphics[width=\figwidthCc]{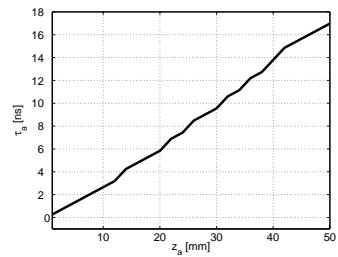}
	\caption{Optimum time-shifts $\tau_a$ as function of the suppression depth $z_a$.}%
	\label{fig:pishift_results}
\end{figure}

The depth-dependent optimum time-shift $\tau_a(z_a)$, with $z_a$ as the chosen depth of increased suppression, is found by maximizing $Q_{z_a}(z_a)$ under variation of $\tau_a$. Fig.~\ref{fig:pishift_results} displays the resulting relation between this depth and the time-shift. 

\subsection{Adjusted synthetic transmit imaging beams}
The $s_-(\vx,t)$ field is adjusted both using the optimum time-shift $\tau_a(z_a)$ and using the general filter $\hh_{z_a}$. The SURF difference HF field is formed for both adjustment methods. %
The depth choice of optimum suppression is varied within $z_a \in[1, 55]$\,mm, and resulting adjusted SURF transmit fields for $z_a = \{5, 10, 20, 30, 40, 55\}$\,mm with $s_-(\vx,t)$ unmodified, modified by the optimum time-shift, and modified by $\hh_{z_a}$, are shown in Figs.~\ref{fig:compare}--\ref{fig:compare_6} displayed at the end of the paper.


\subsection{Beam quality ratios}
\begin{figure}[!tb]
	\centering
	\includegraphics[width=\figwidthCc]{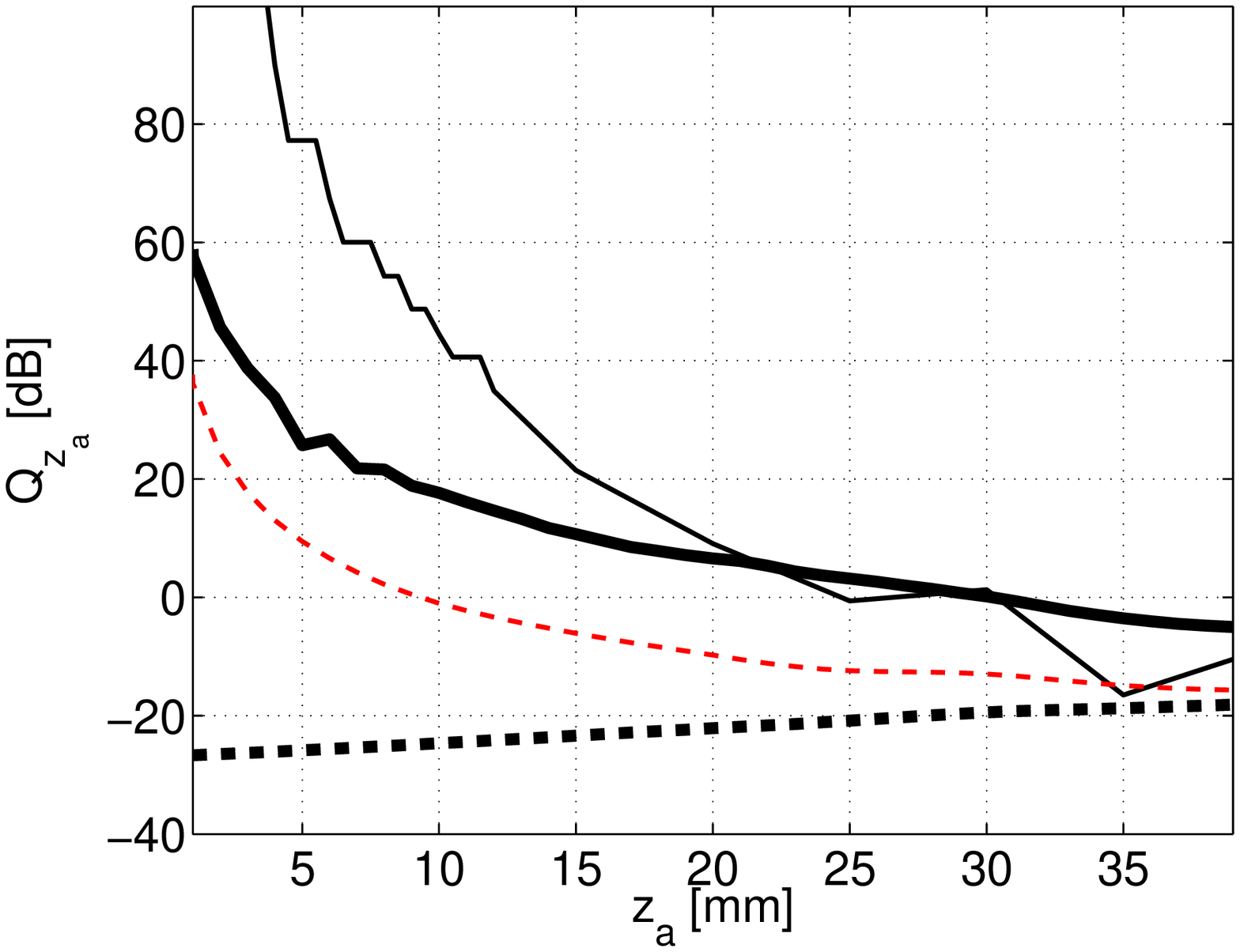} %
	\caption[Specific beam quality ratio.]{Specific beam quality $Q_{z_a}$.\includegraphics{thick_line}: $\tau_a$ adjusted,\includegraphics{thin_line}: $\hh_{z_a}$ adjusted,\includegraphics{thin_dashed_line_red}: non-adjusted,\includegraphics{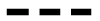}: HF beam without LF manipulation (fundamental imaging).}
	\label{fig:Qza}%
\end{figure}

Fig.~\ref{fig:Qza} shows a comparison between the specific beam quality ratio $Q_{z_a}(z_a)$ calculated for beams with different adjustments, for the non-adjusted beam, and for the conventional fundamental imaging HF transmit-beam without LF manipulation.
\section{Discussion\label{sec:discussion}}

\begin{figure*}[!p]
	\centering
	\subfigure[SURF pressure beams. \panedescription]{\makebox[\figw][c]{\includegraphics[width=\figw]{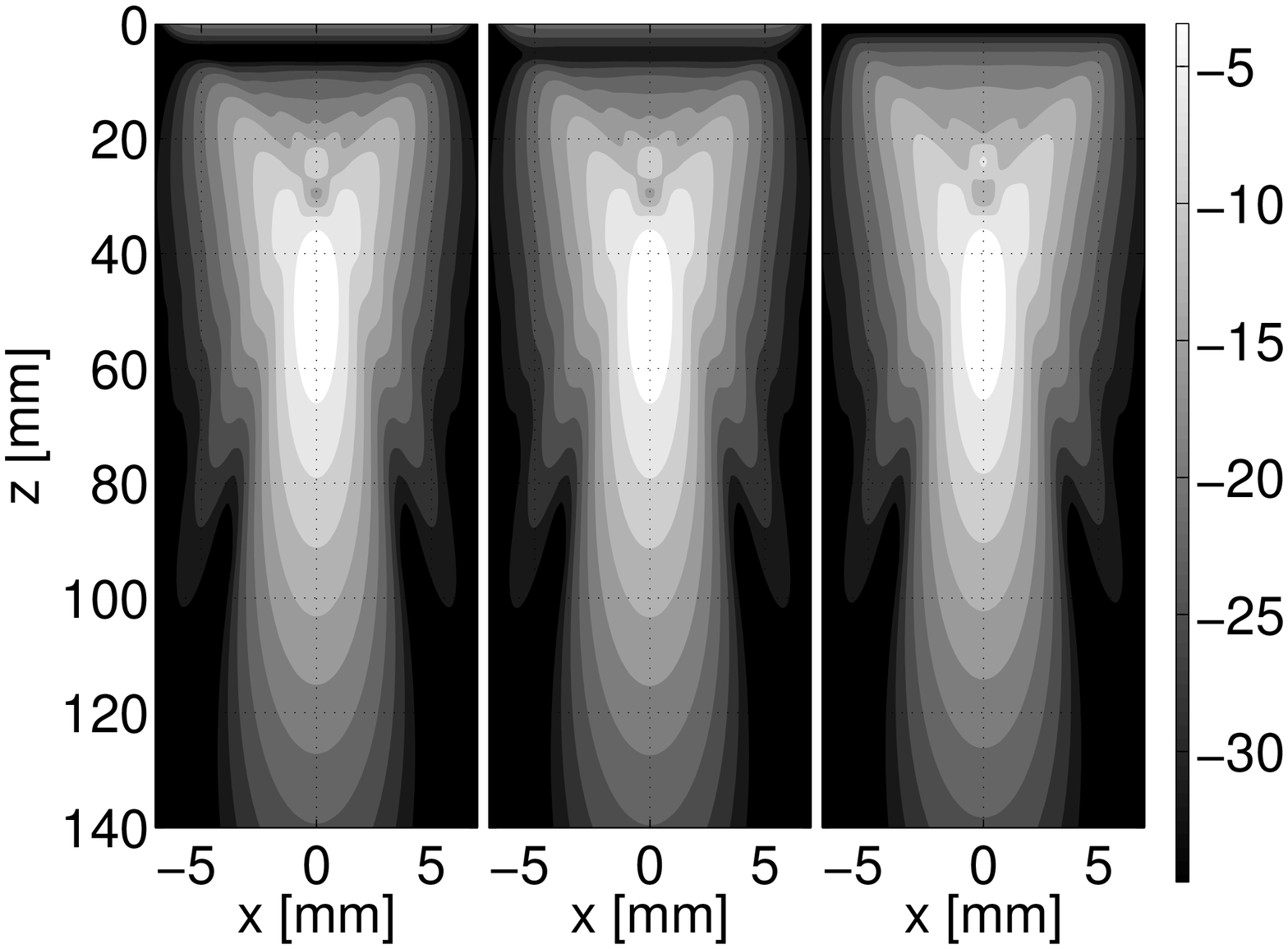}}}%
	\quad\quad
	\subfigure[On-axis max. SURF HF difference pressure. Top: individual normalization, bottom: common norm. \label{subfig:b}]{\makebox[\figw][c]{\includegraphics[width=\figw]{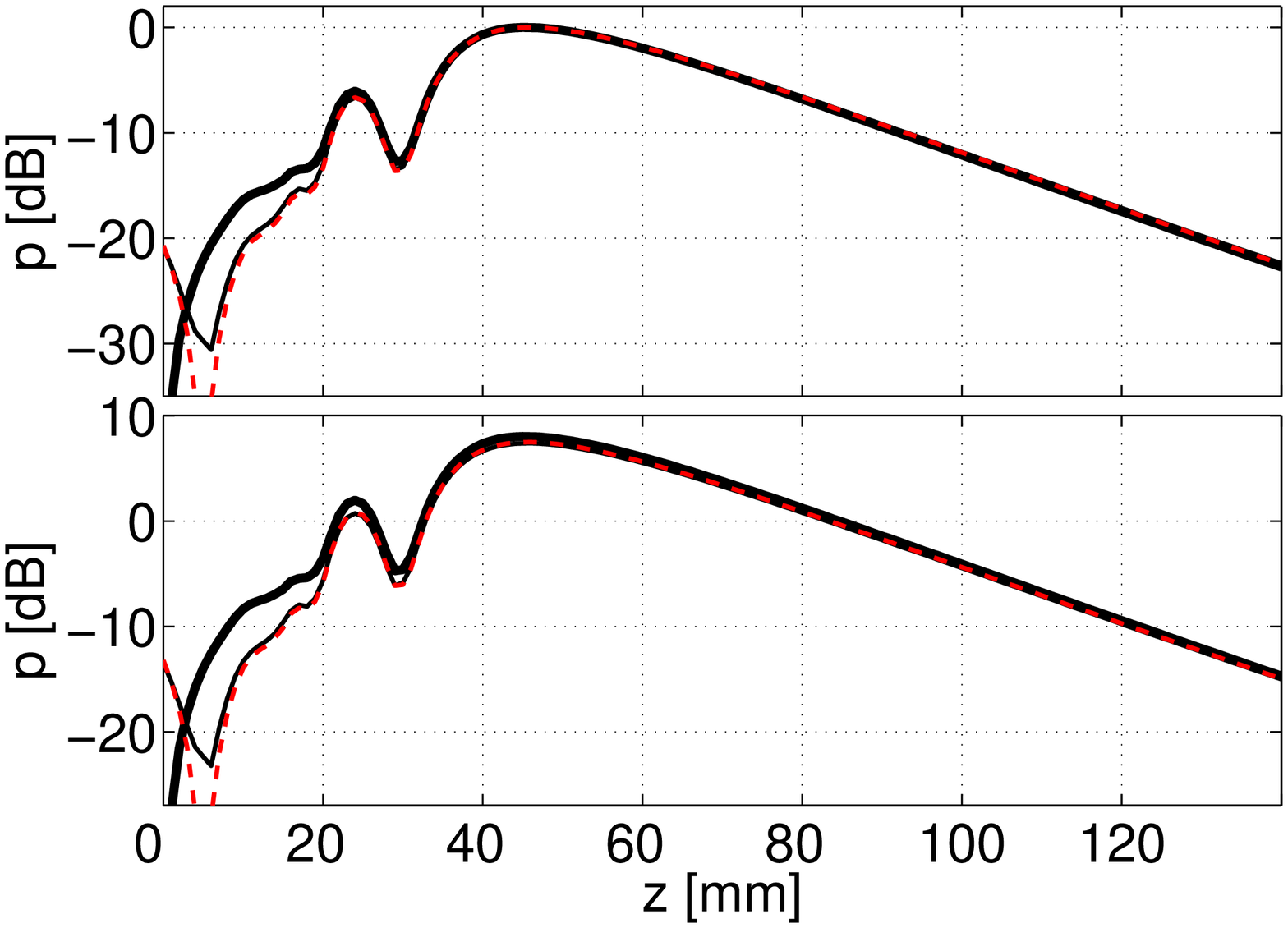}}}\\%
	\subfigure[On-axis SURF pulses][On-axis SURF pulses. \panedescription]{\makebox[\figw][c]{\includegraphics[width=\figw]{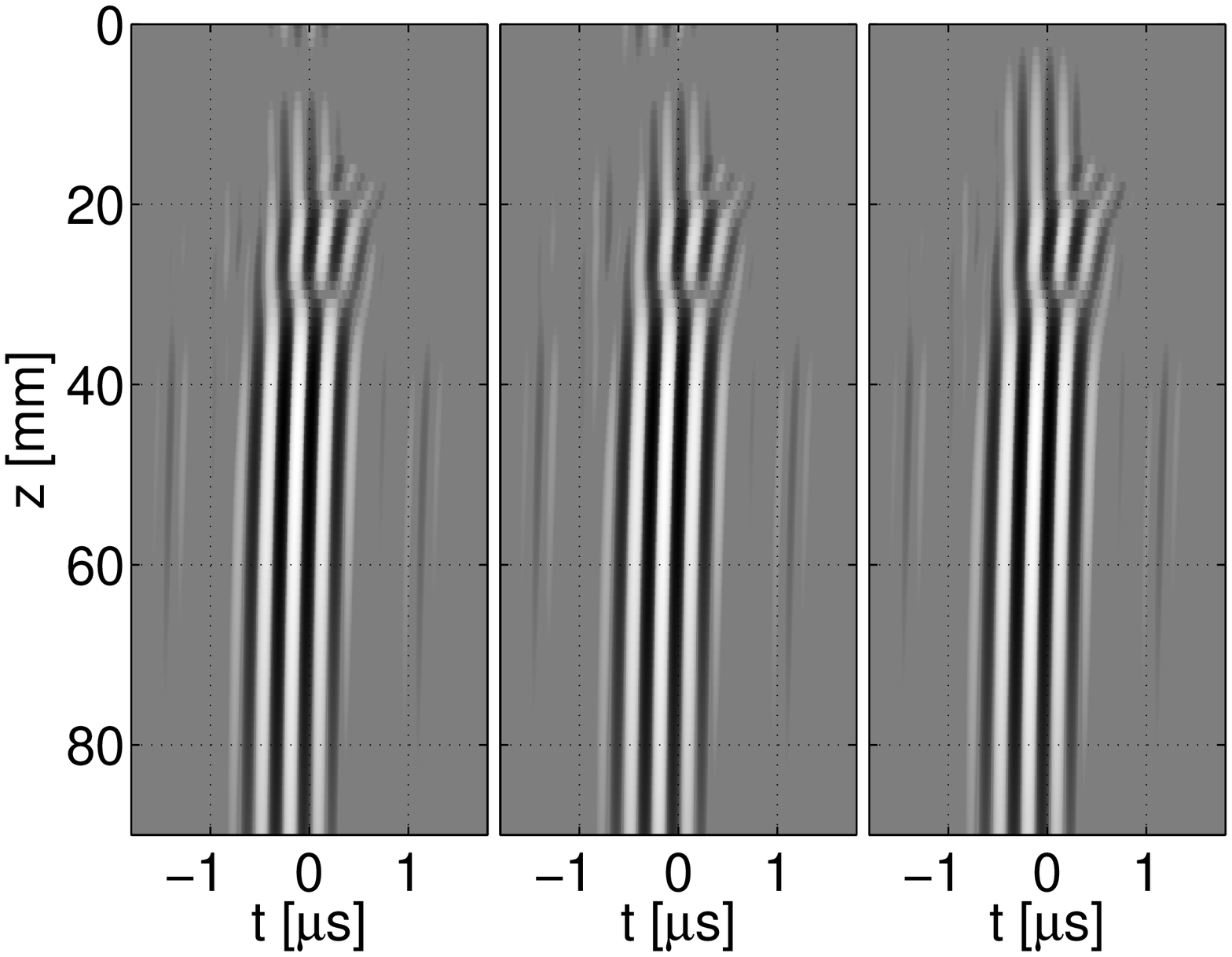}}}%
	\quad\quad
	\subfigure[On-axis SURF pulses. Top pane: at the depth $z_a$, bottom: at the focal depth. \label{subfig:dC}]{\makebox[\figw][c]{\includegraphics[width=\figw]{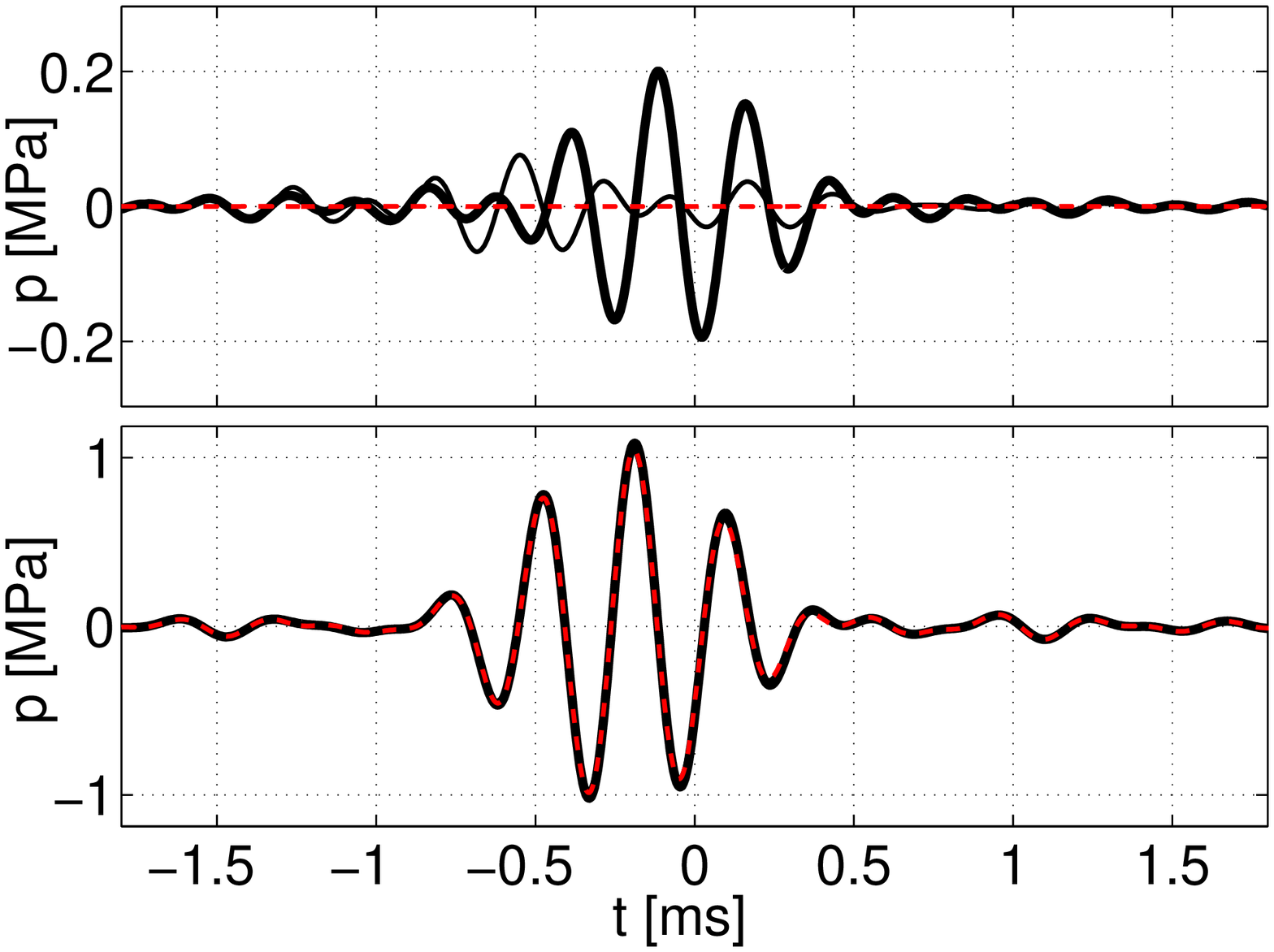}}}%
	\caption[SURF pulses and fields.]{Transmit beam and on-axis pulse comparisons, \fbox{$z_a=5$\,mm.} SURF pulses and fields generated from simulated data. Line notation for \subref{subfig:b} and \subref{subfig:dC}: \linedescription.}%
	\label{fig:compare}%
\end{figure*}

\begin{figure*}[!p]
	\centering
	\begin{tabular}{c|c}
	\includegraphics[width=\figw]{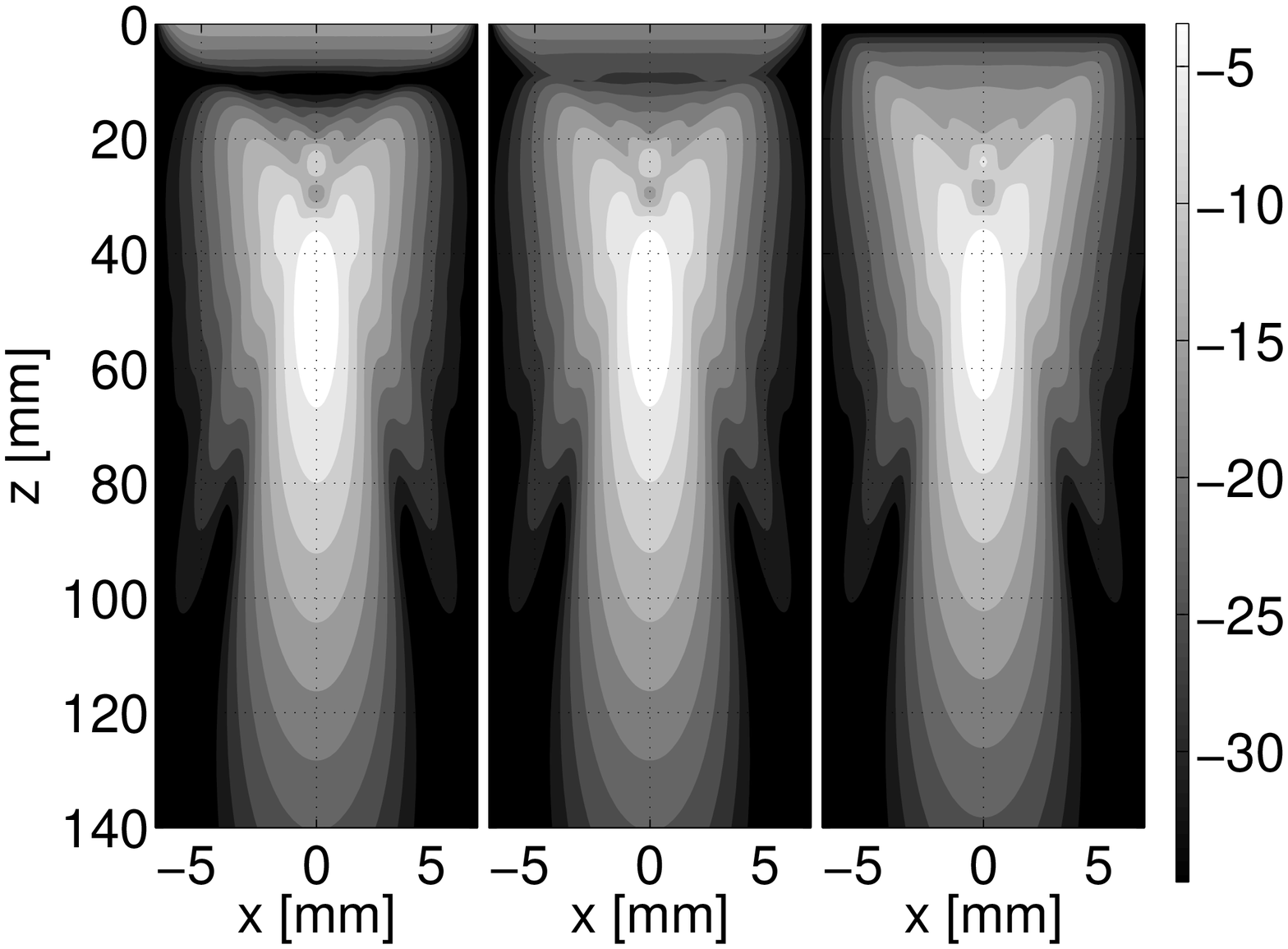}%
	&
	\includegraphics[width=\figw]{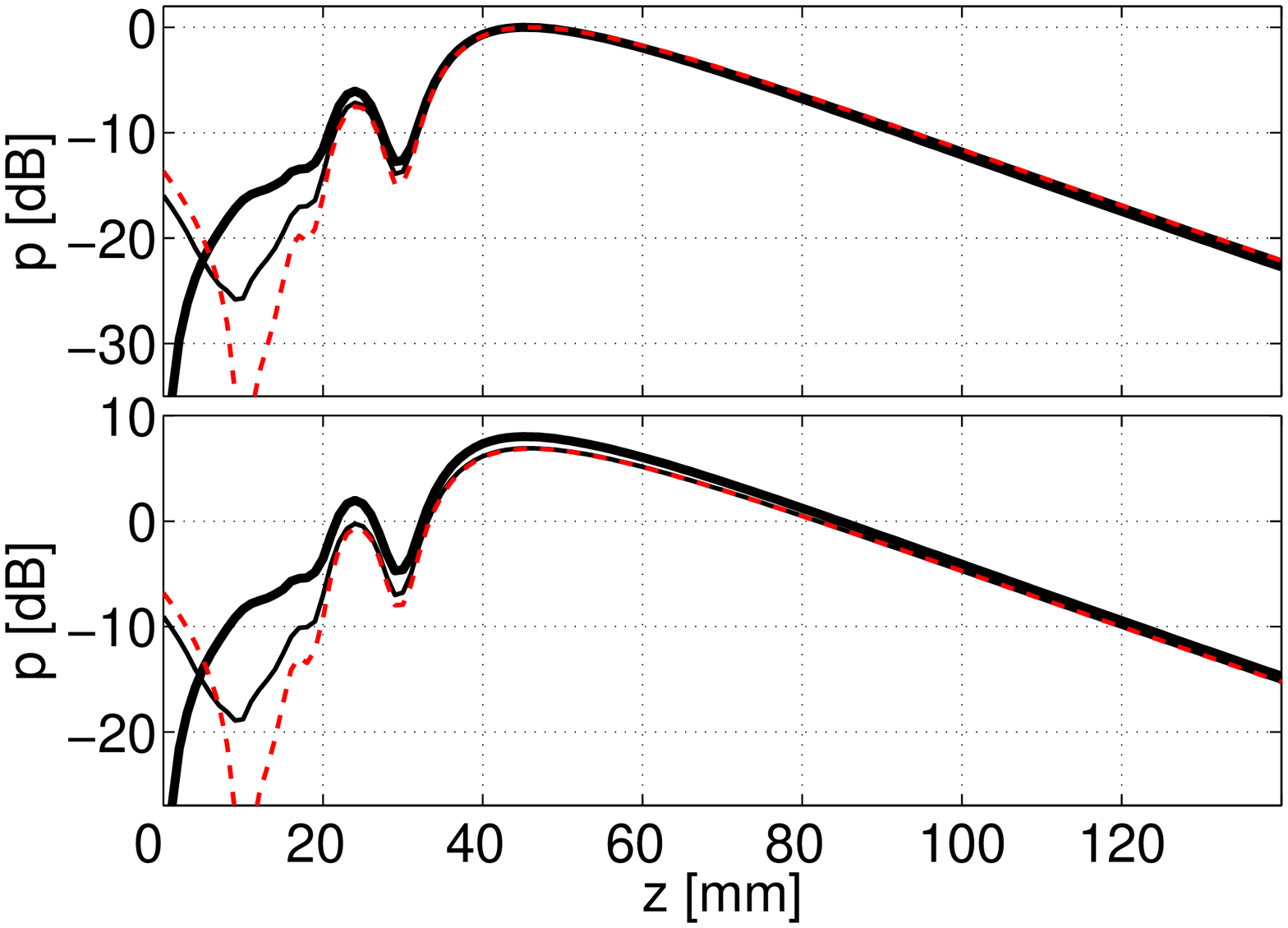}\\%
	\hline
	\includegraphics[width=\figw]{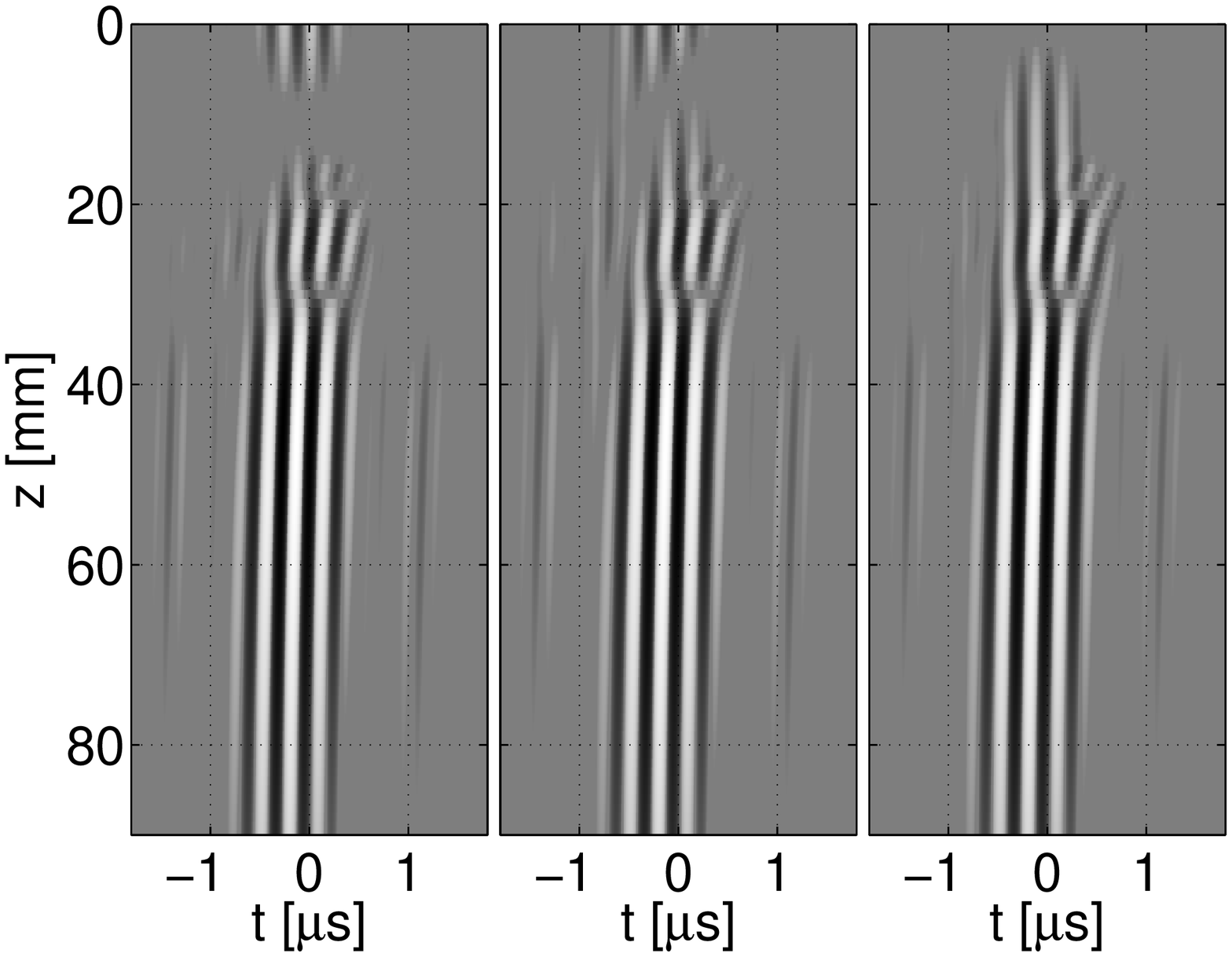}%
	&
	\includegraphics[width=\figw]{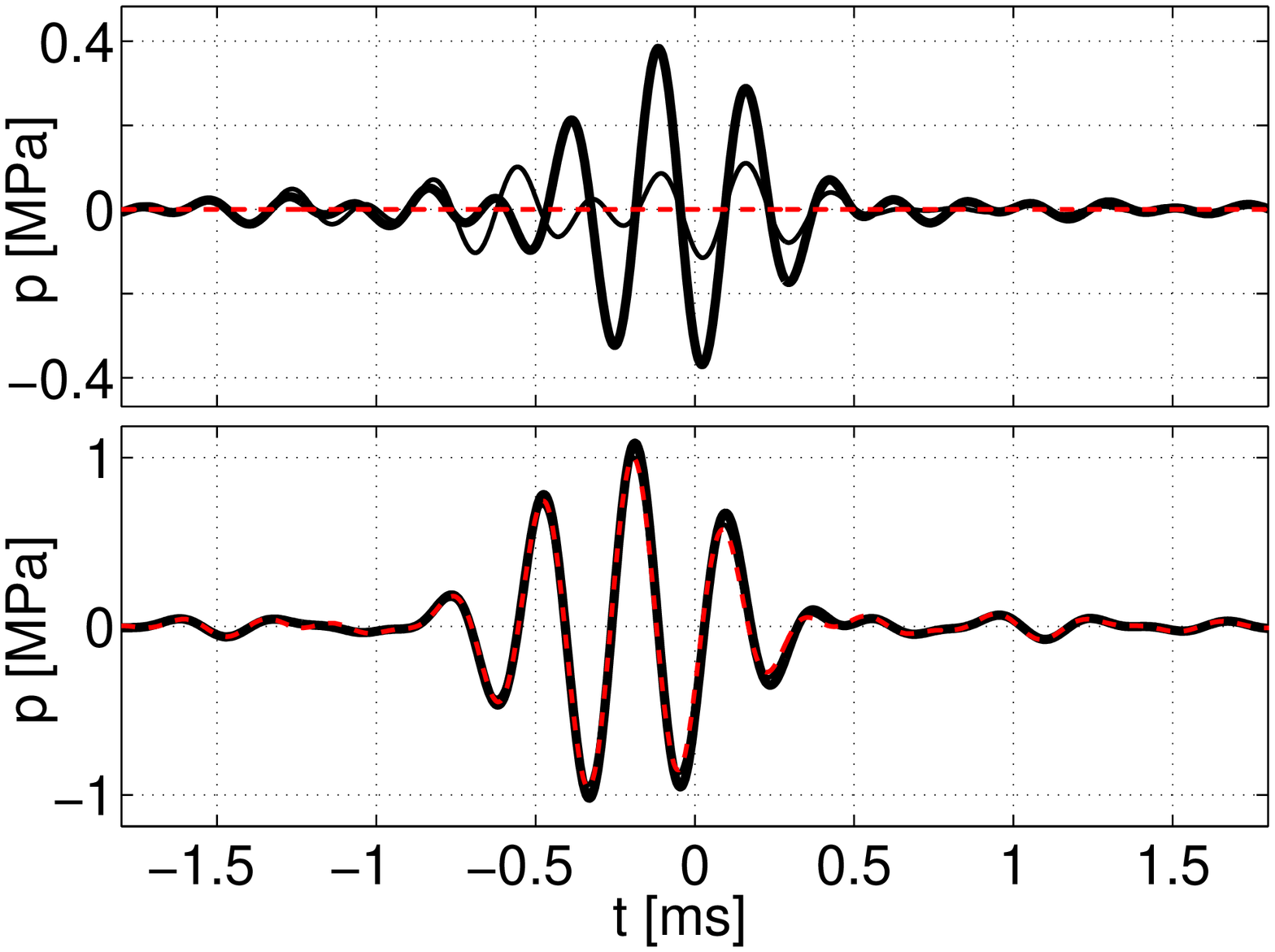}%
	\end{tabular}
	\caption[SURF pulses and fields.]{Transmit beam and on-axis pulse comparisons, \fbox{$z_a=10$\,mm.} Notation as in Fig.~\ref{fig:compare}.}
	\label{fig:compare_2}%
\end{figure*}

\begin{figure*}[!p]
	\centering
	\begin{tabular}{c|c}
	\includegraphics[width=\figw]{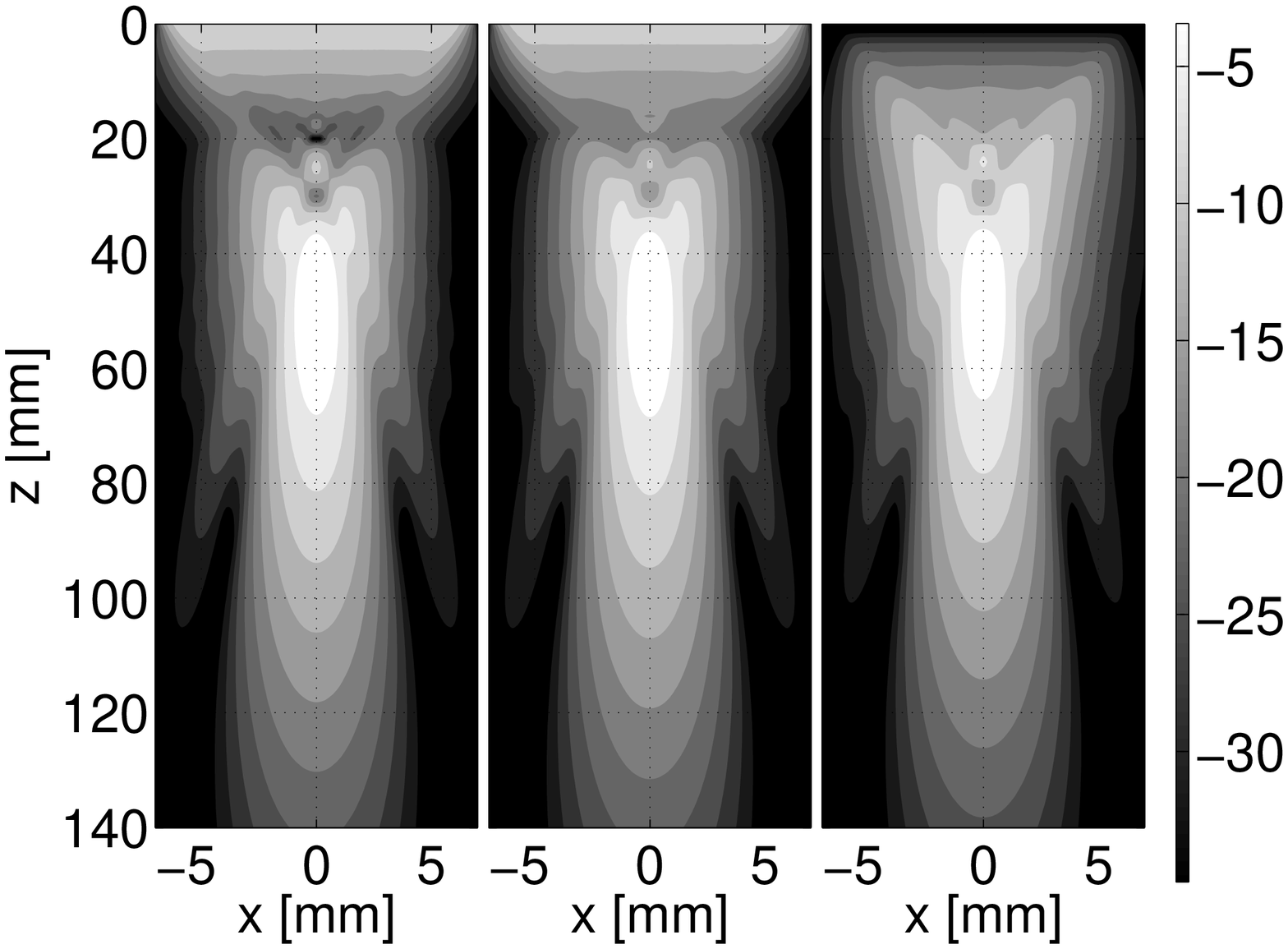}%
	&
	\includegraphics[width=\figw]{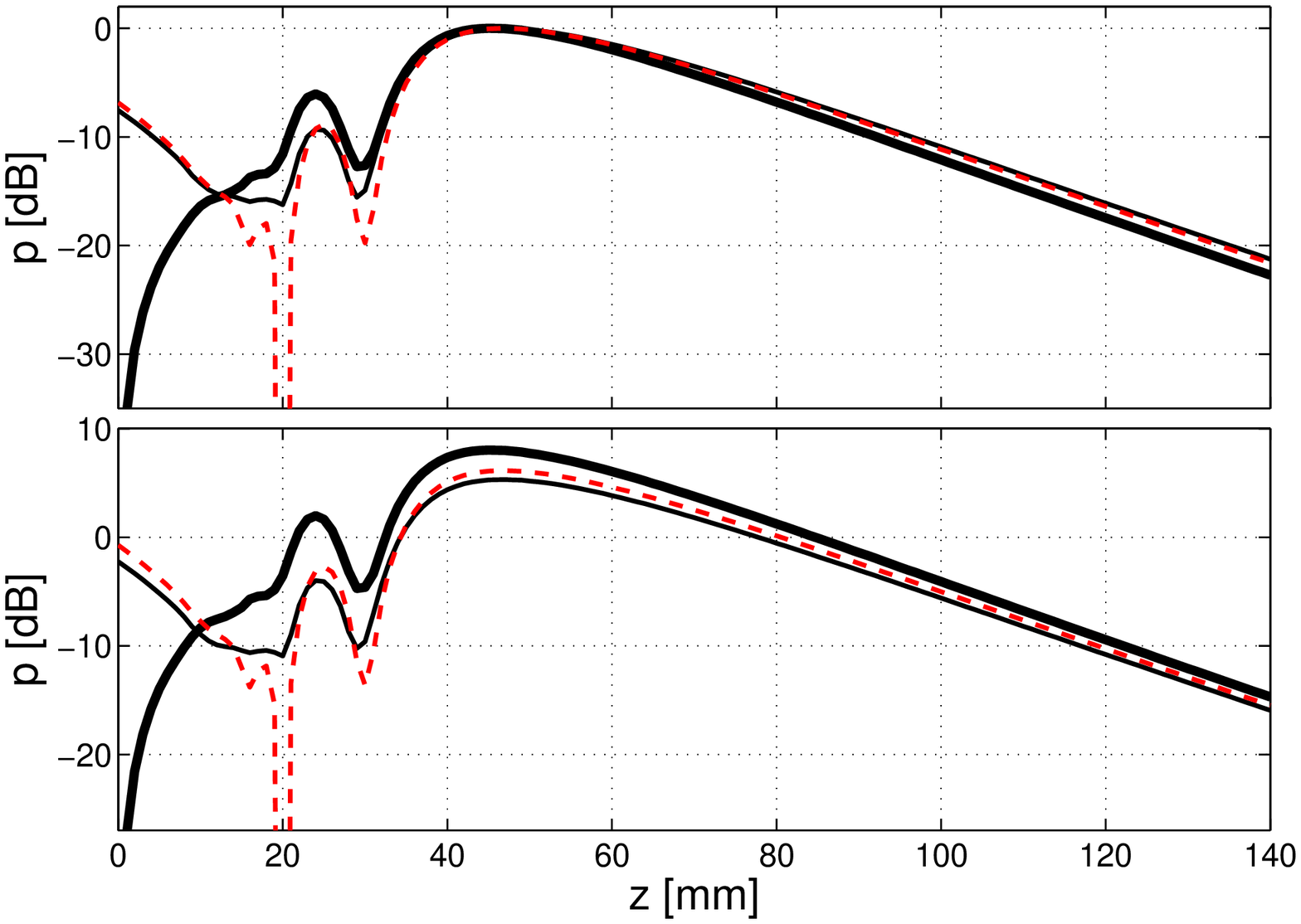}\\%
	\hline
	\includegraphics[width=\figw]{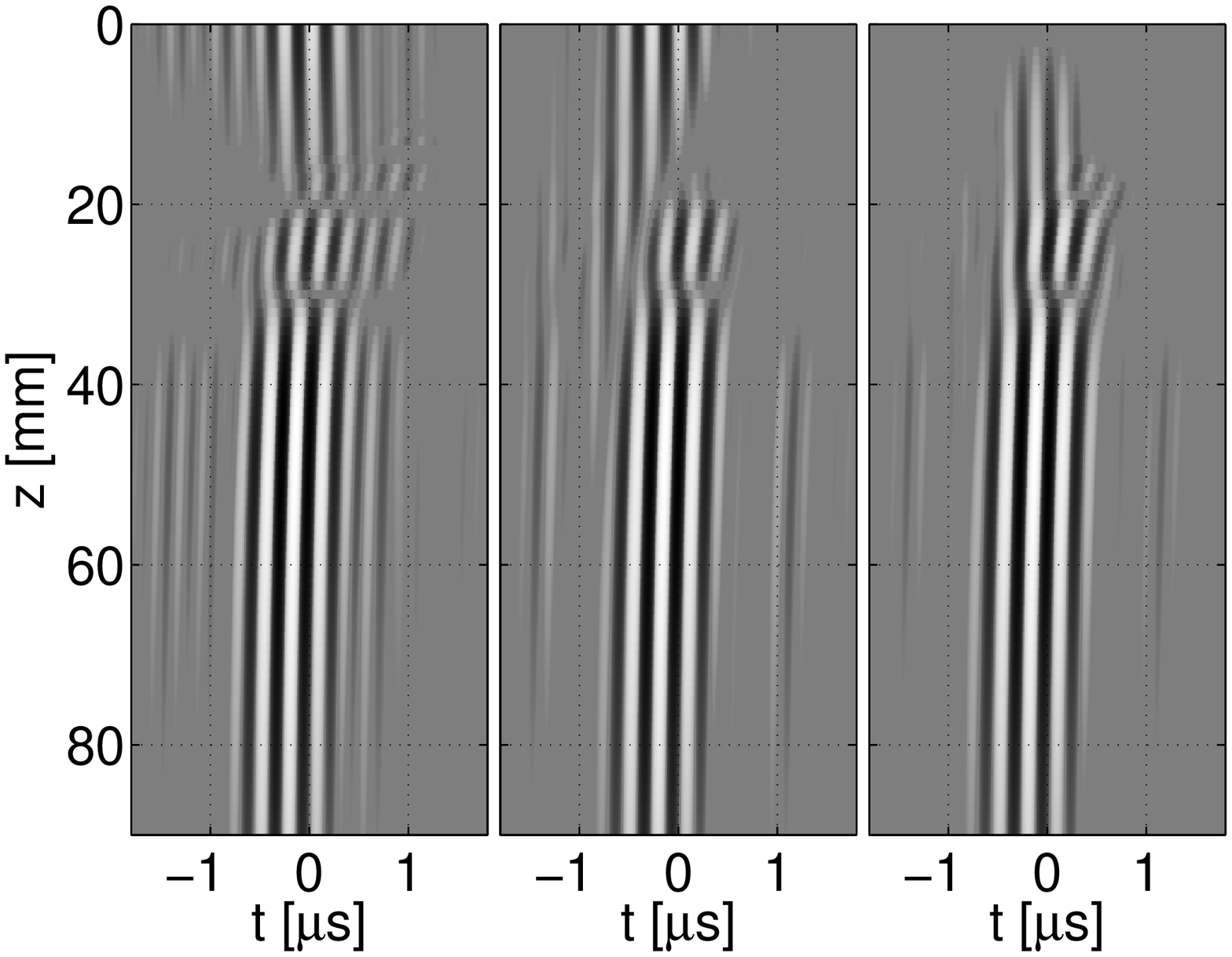}%
	&
	\includegraphics[width=\figw]{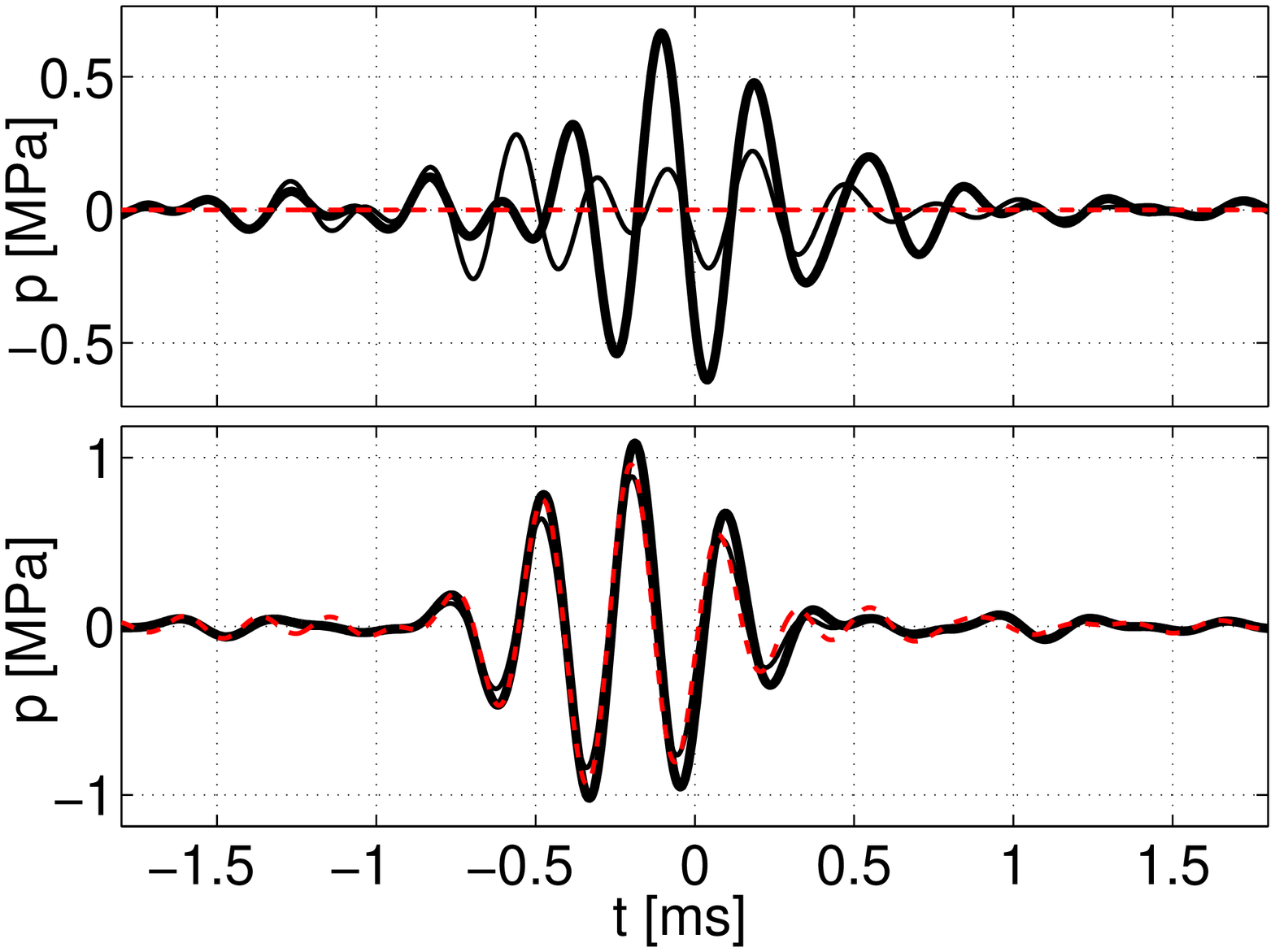}%
	\end{tabular}
	\caption[SURF pulses and fields.]{Transmit beam and on-axis pulse comparisons, \fbox{$z_a=20$\,mm.} Notation as in Fig.~\ref{fig:compare}.}
	\label{fig:compare_3}%
\end{figure*}

\begin{figure*}[!p]
	\centering
	\begin{tabular}{c|c}
	\includegraphics[width=\figw]{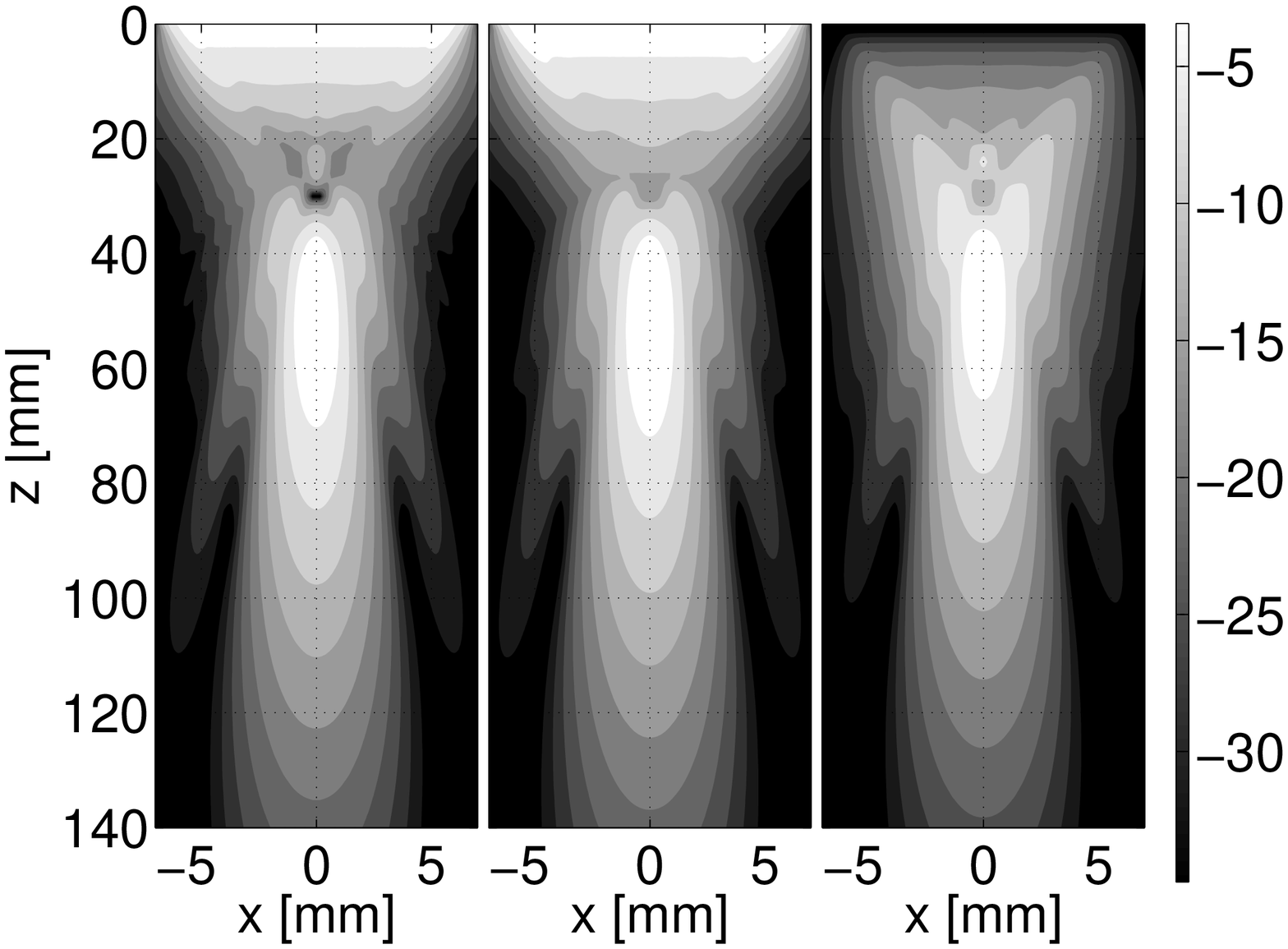}%
	&
	\includegraphics[width=\figw]{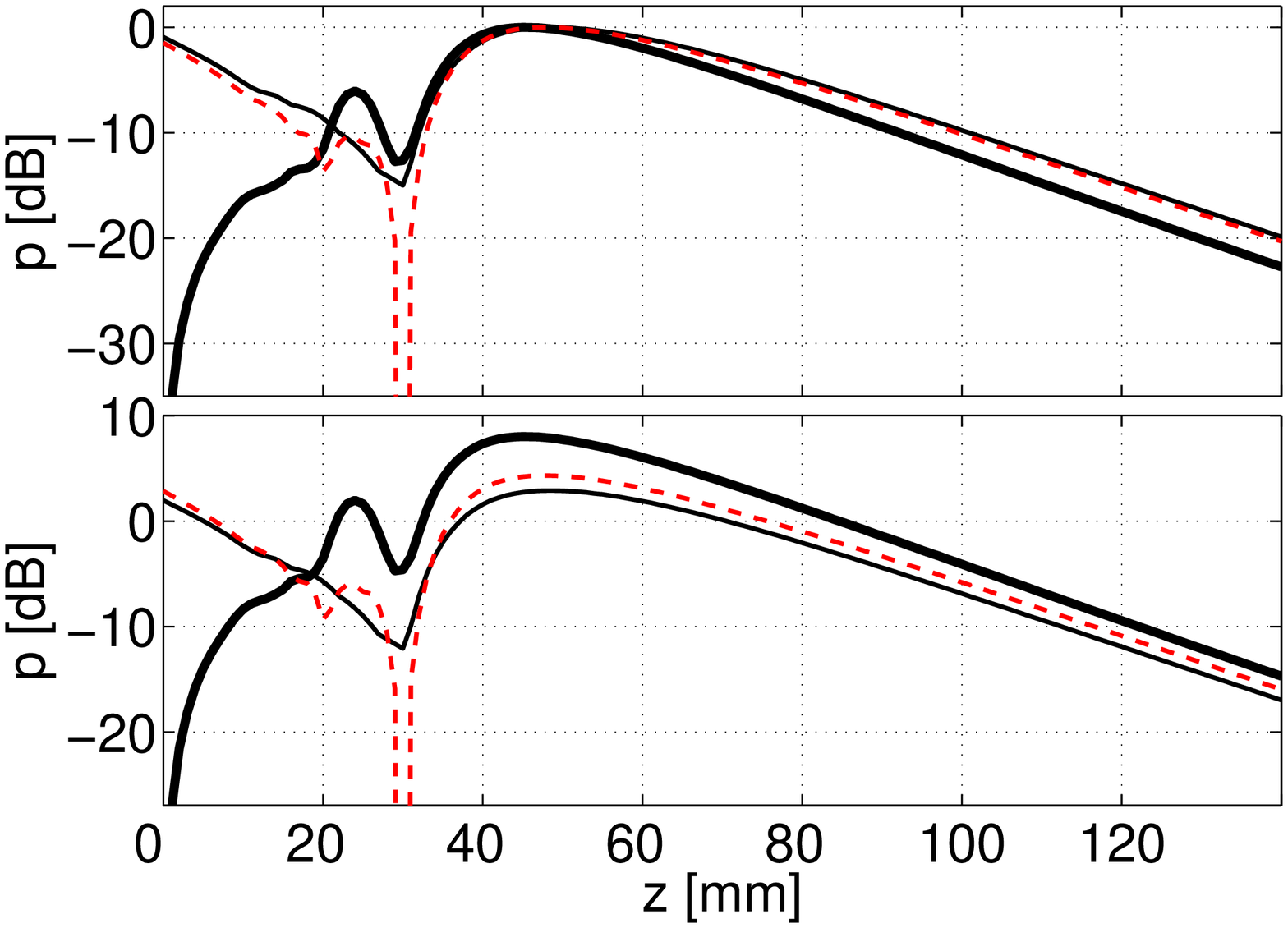}\\%
	\hline
	\includegraphics[width=\figw]{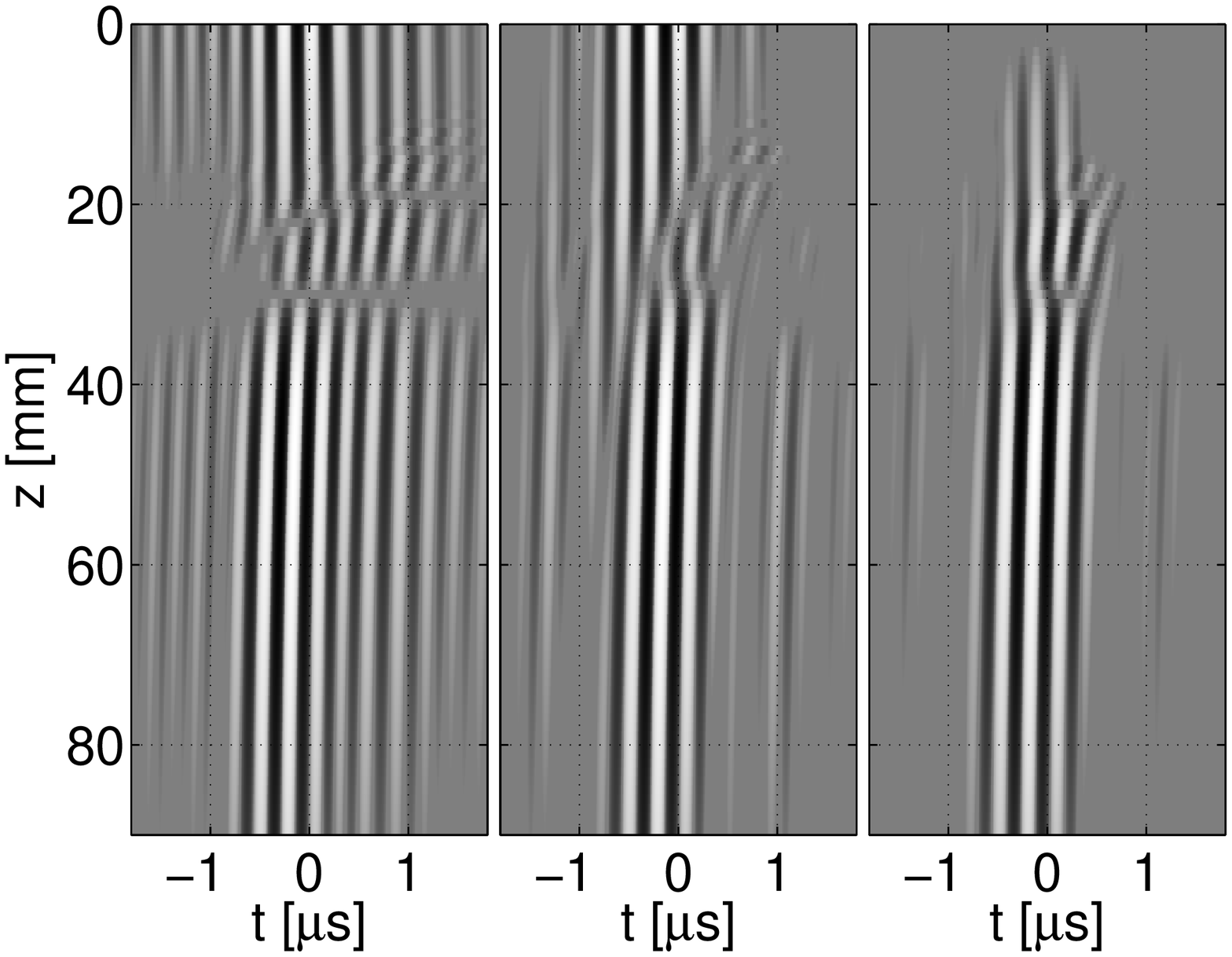}%
	&
	\includegraphics[width=\figw]{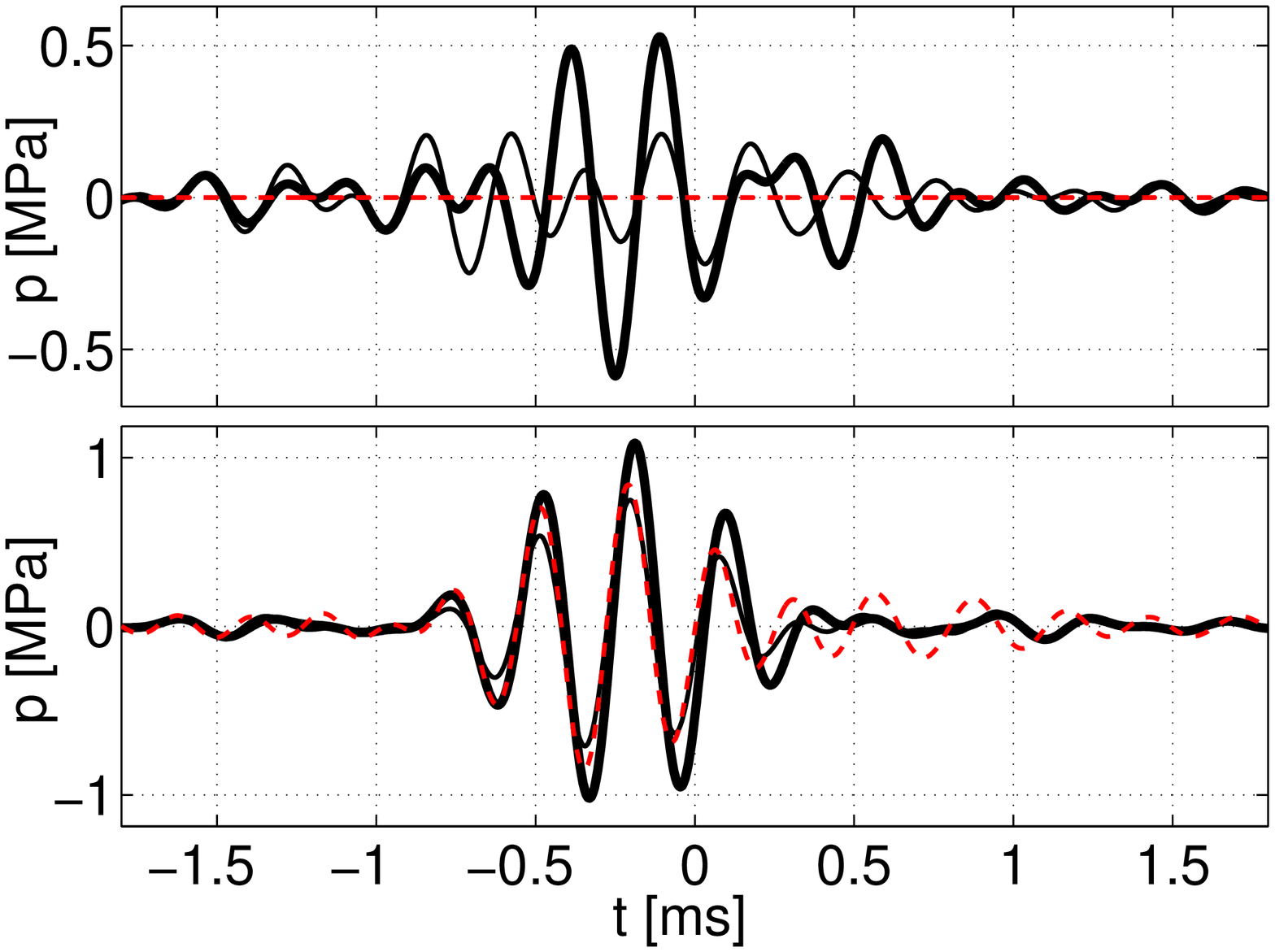}%
	\end{tabular}
	\caption[SURF pulses and fields.]{Transmit beam and on-axis pulse comparisons, \fbox{$z_a=30$\,mm.} Notation as in Fig.~\ref{fig:compare}.}
	\label{subfig:bad_focalfield}%
\end{figure*}

\begin{figure*}[!p]
	\centering
	\begin{tabular}{c|c}
	\includegraphics[width=\figw]{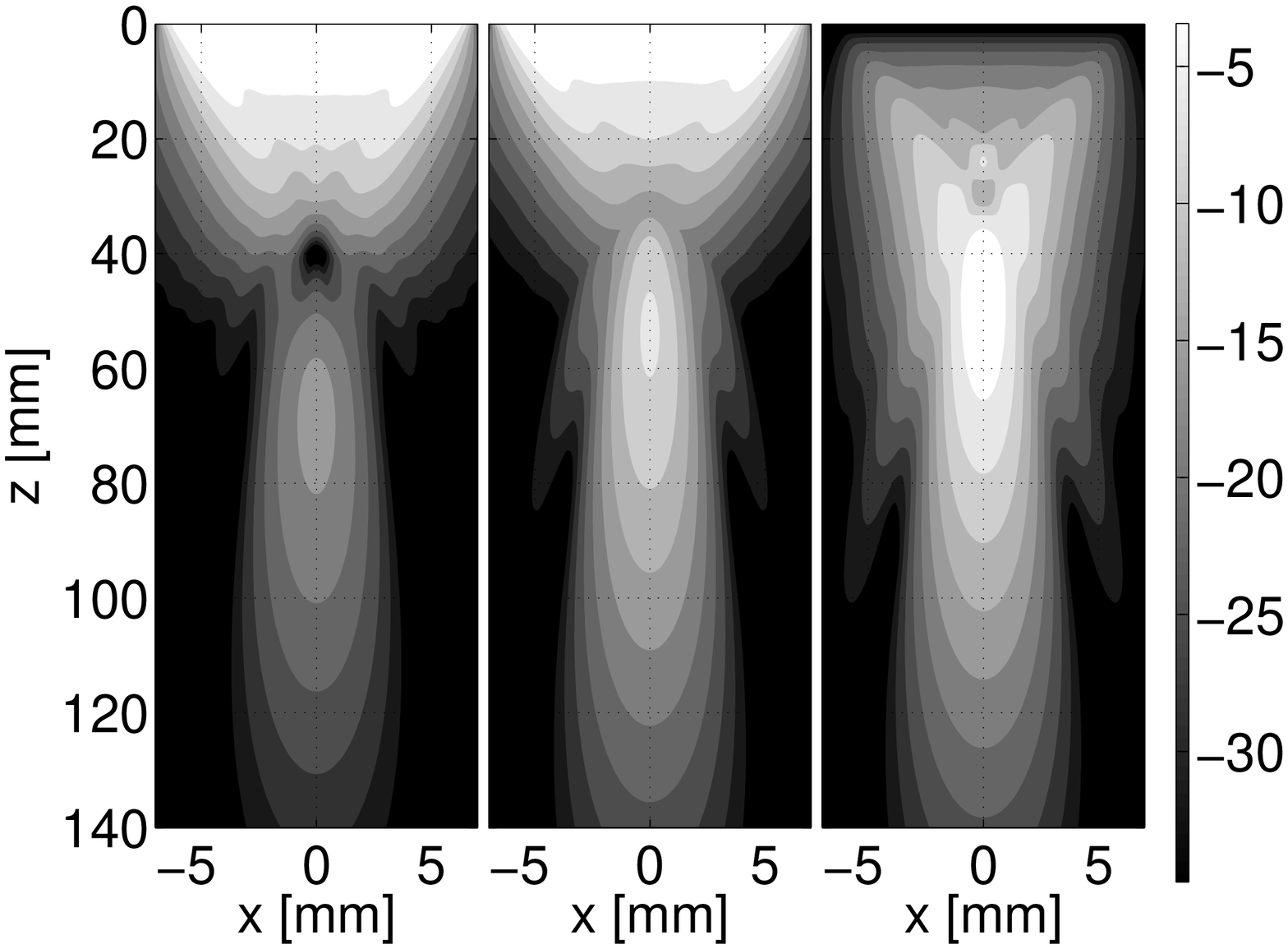}%
	&
	\includegraphics[width=\figw]{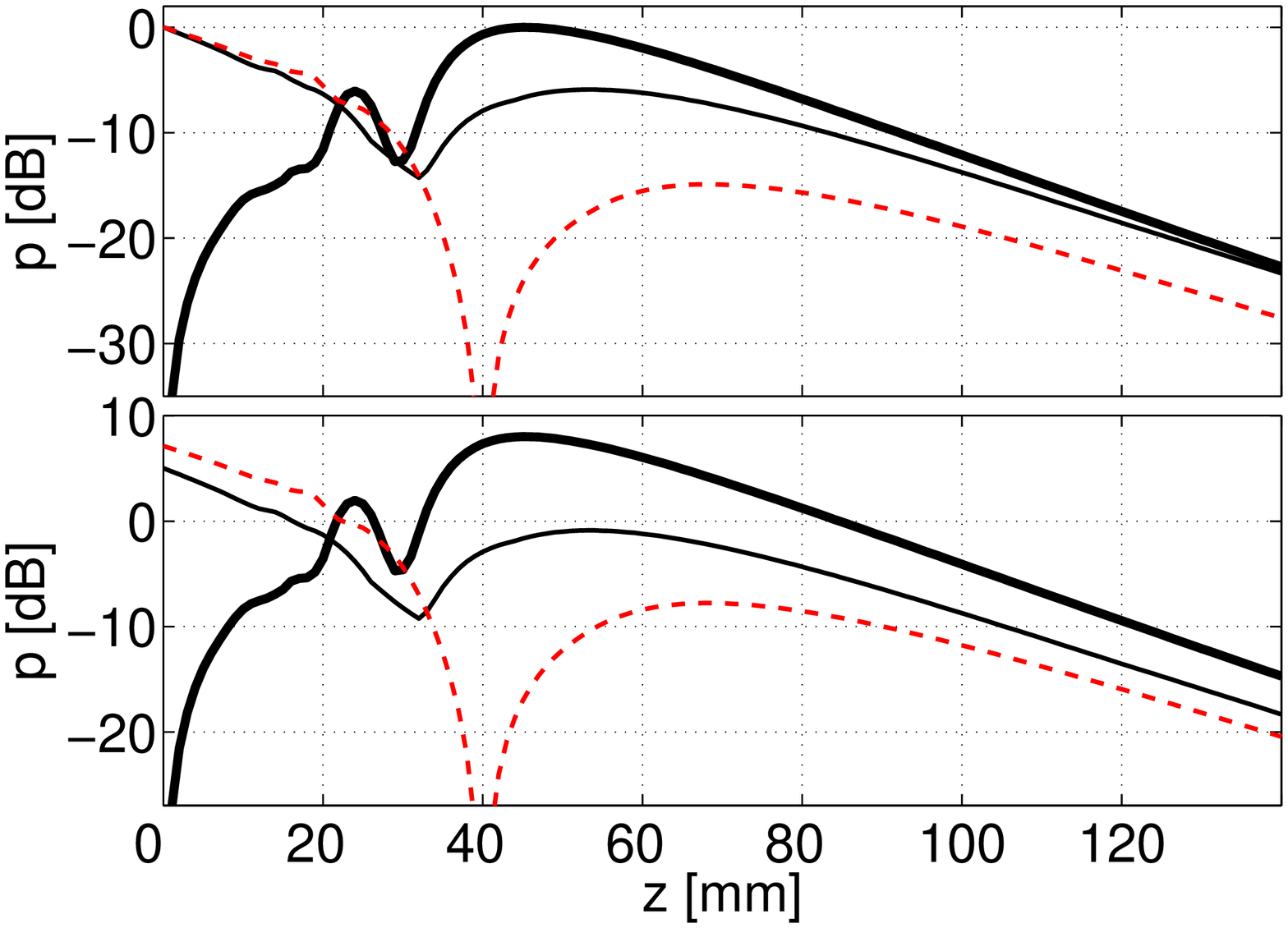}\\%
	\hline
	\includegraphics[width=\figw]{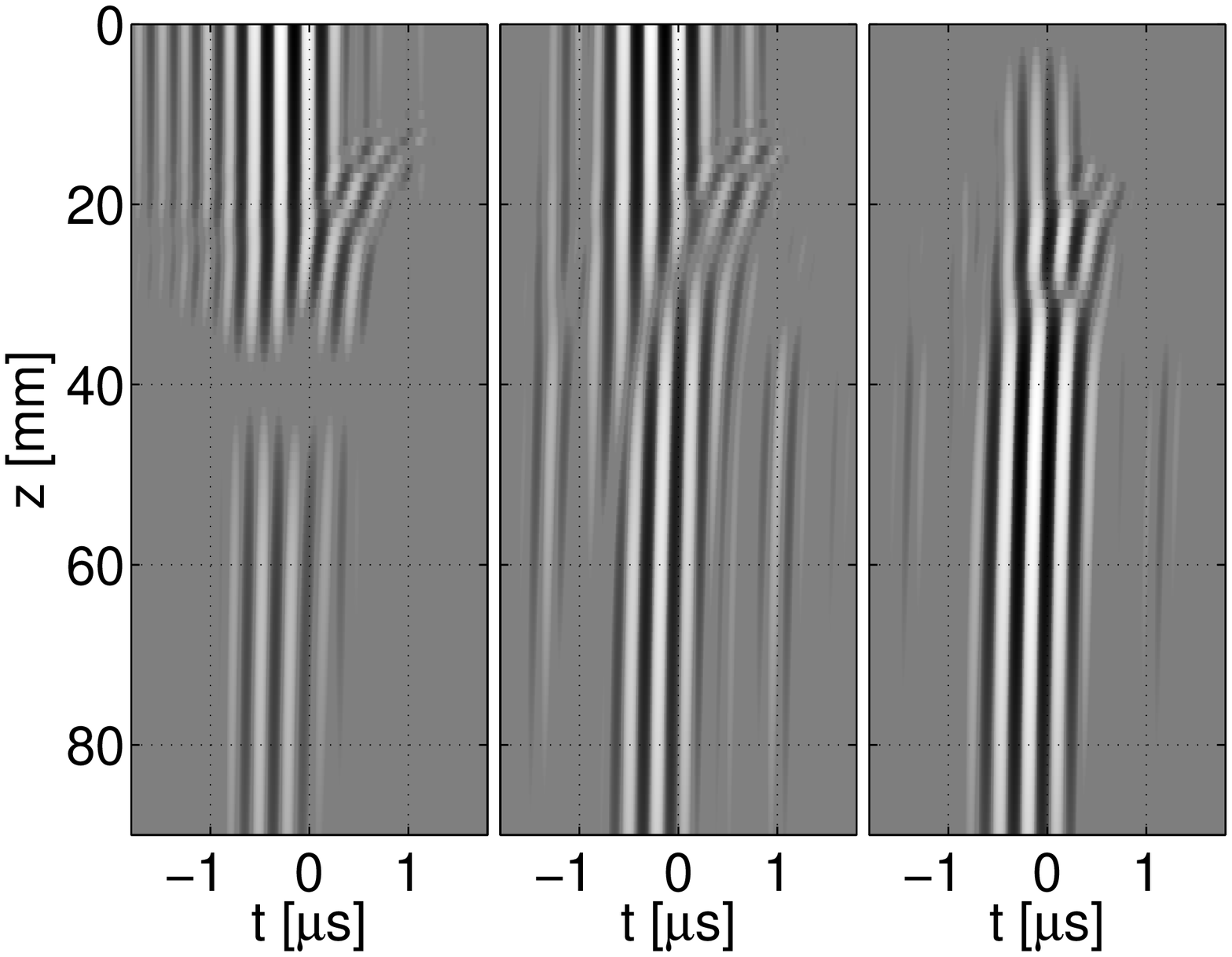}%
	&
	\includegraphics[width=\figw]{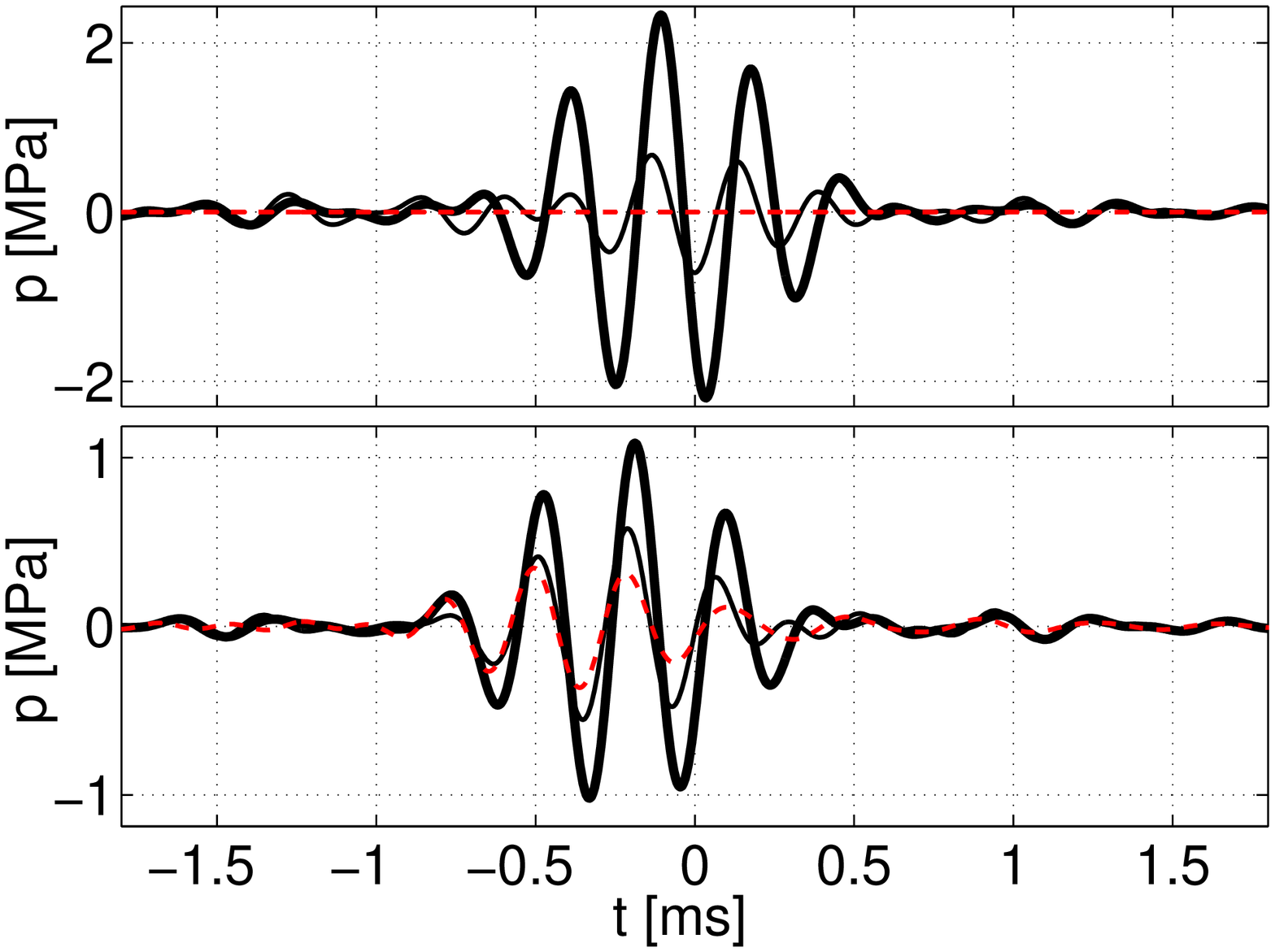}%
	\end{tabular}
	\caption[SURF pulses and fields.]{Transmit beam and on-axis pulse comparisons, \fbox{$z_a=40$\,mm.} Notation as in Fig.~\ref{fig:compare}.}
	\label{fig:compare_5}%
\end{figure*}

\begin{figure*}[!p]
	\centering
	\begin{tabular}{c|c}
	\includegraphics[width=\figw]{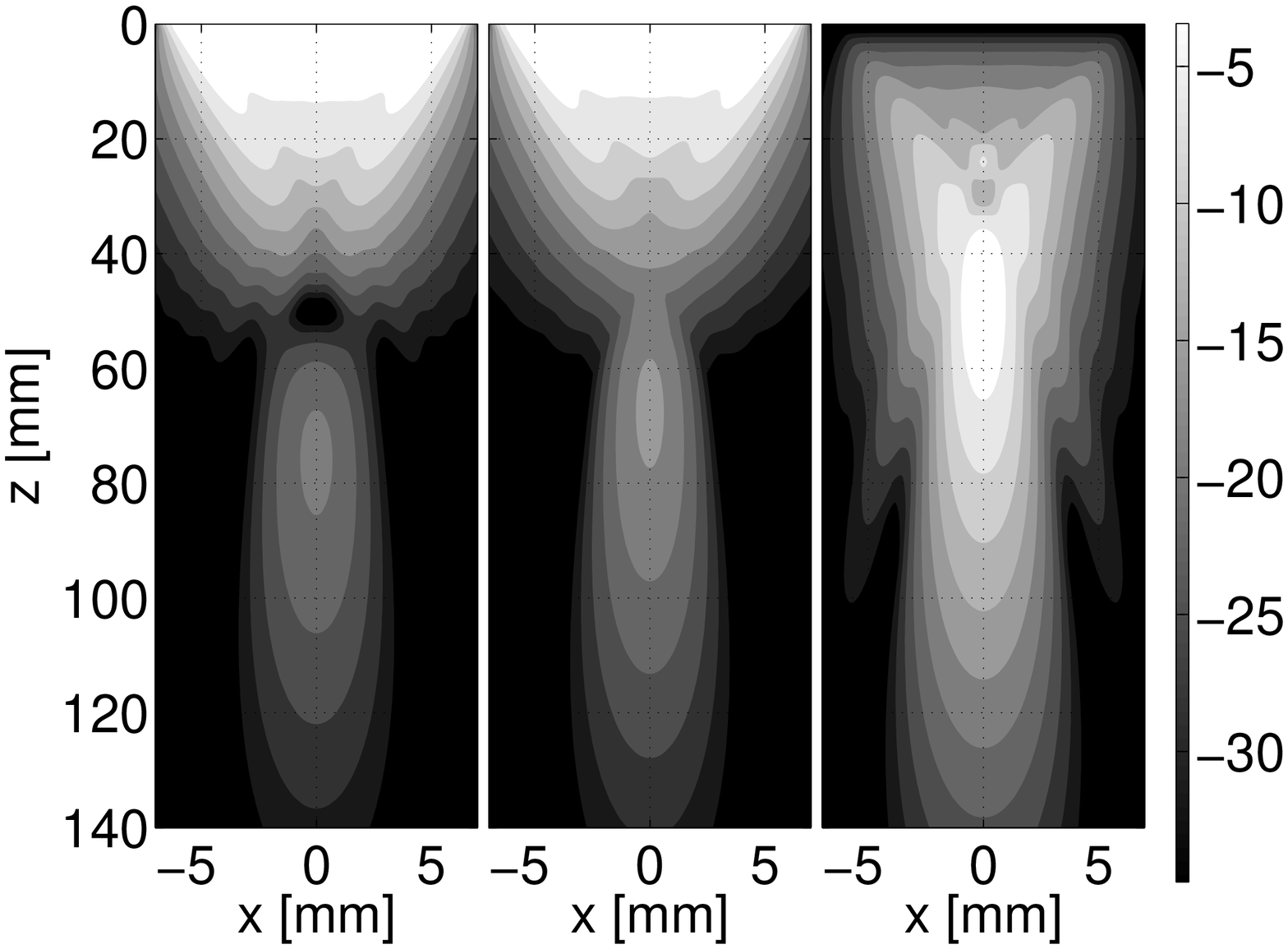} %
	&
	\includegraphics[width=\figw]{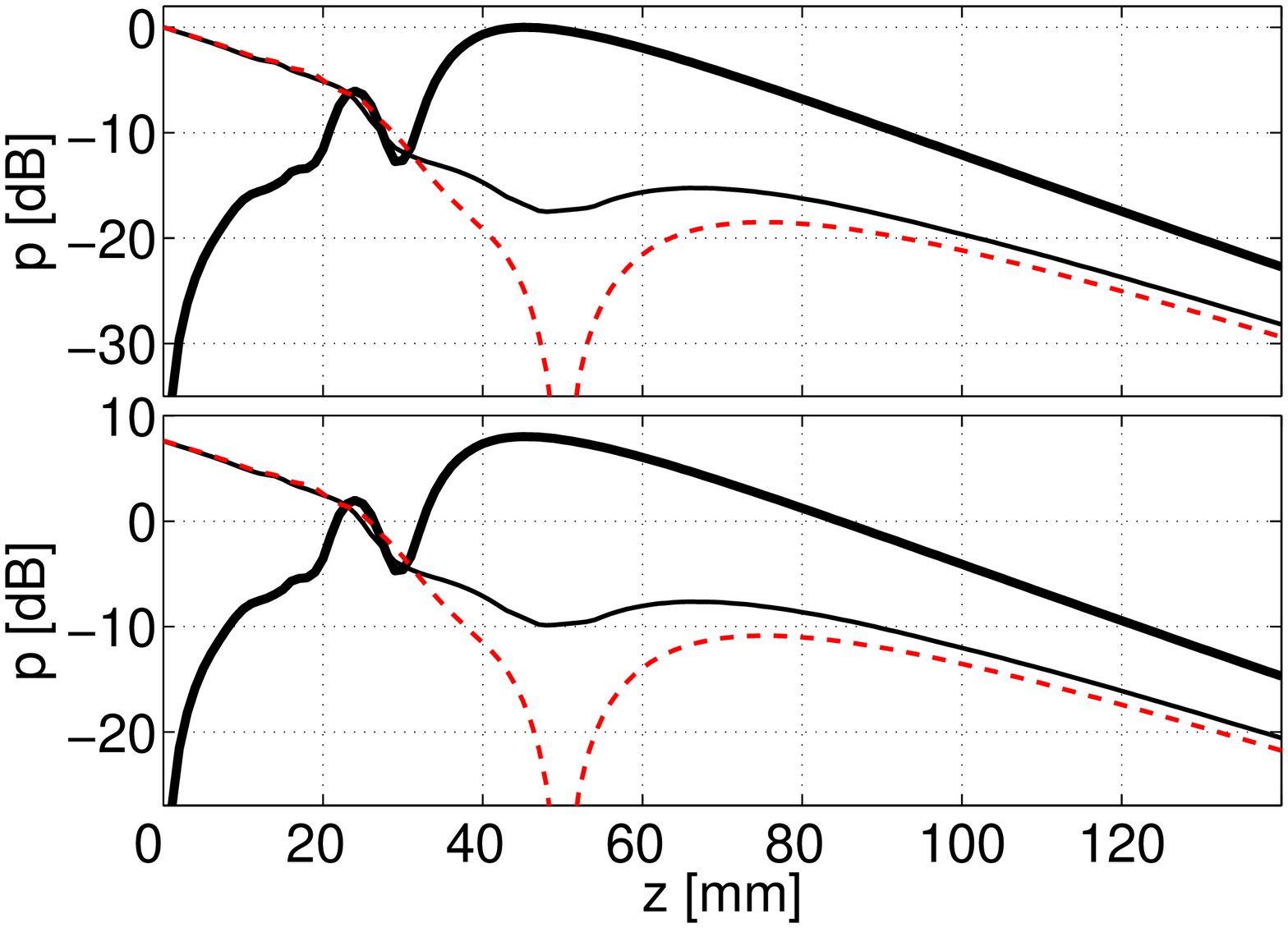}\\ %
	\hline
	\includegraphics[width=\figw]{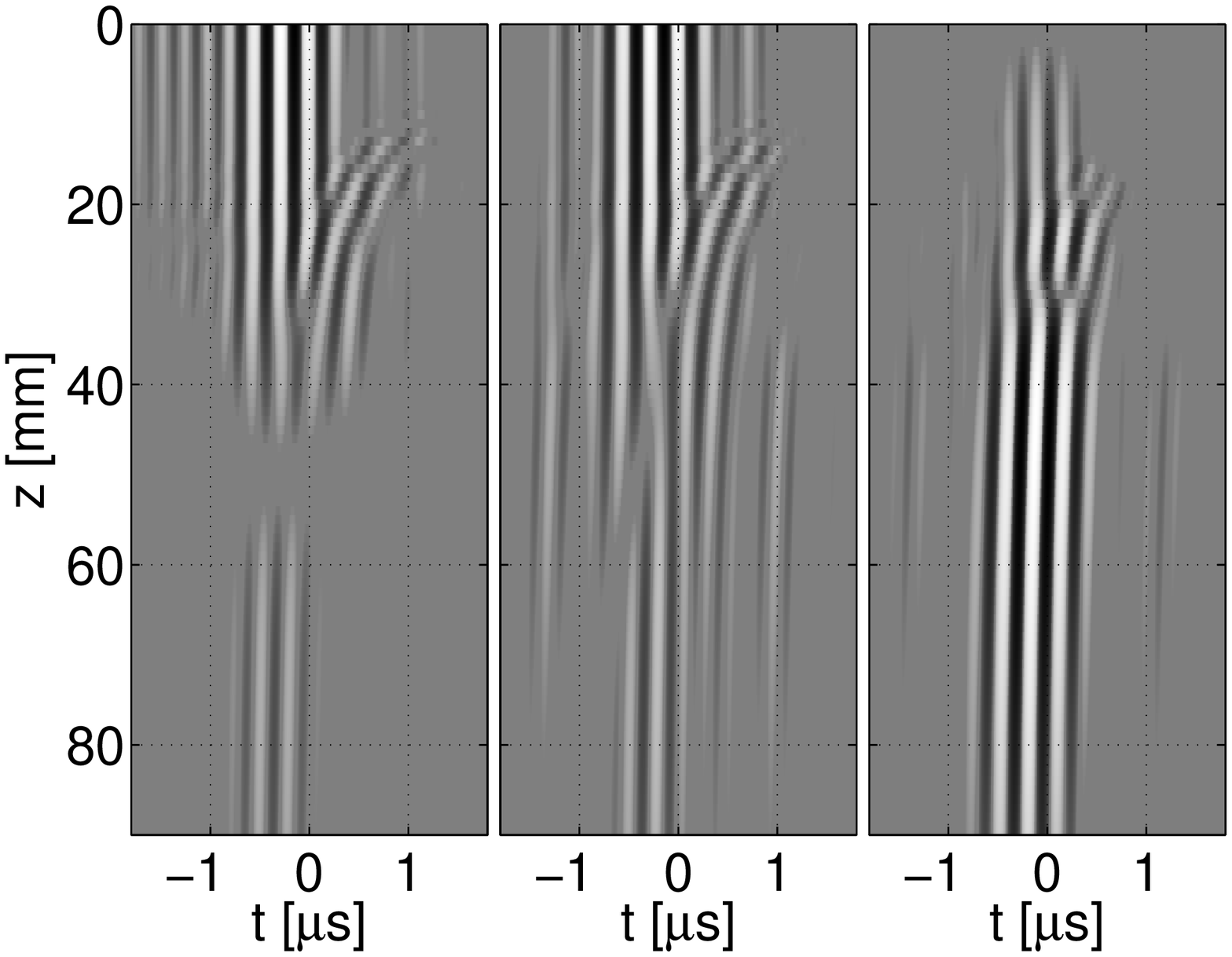} %
	&
	\includegraphics[width=\figw]{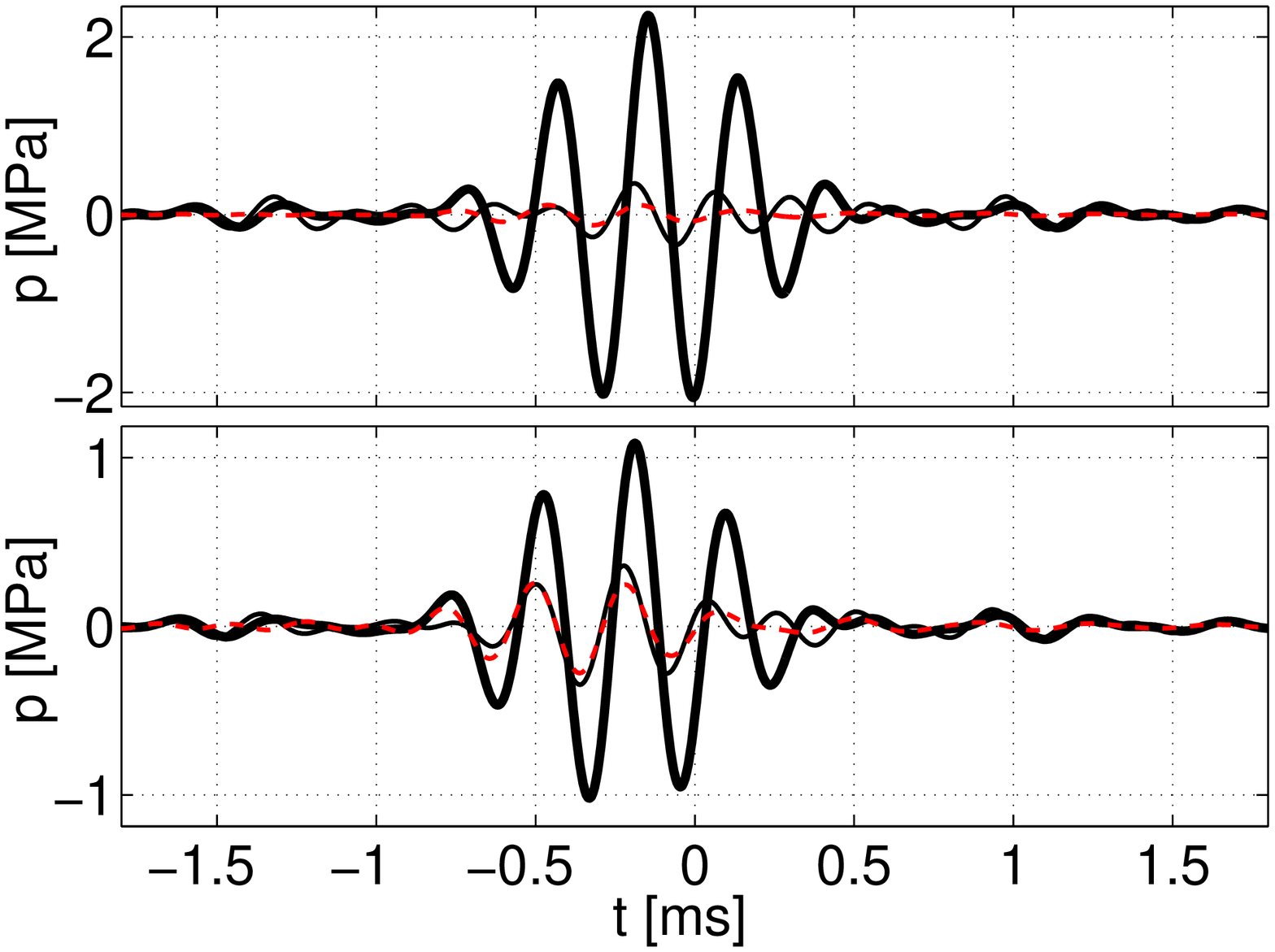}%
	\end{tabular}
	\caption[SURF pulses and fields.]{Transmit beam and on-axis pulse comparisons, \fbox{$z_a=55$\,mm.} Notation as in Fig.~\ref{fig:compare}.}
	\label{fig:compare_6}%
\end{figure*}

As illustrated in Figs.~\ref{fig:compare}--\ref{subfig:bad_focalfield}, the realized time-shift adjustments on the simulated data generate a decrease of on-axis amplitude of the individually normalized SURF beams of ${\sim}\{10, 9, 4, 3\}$\,dB at the chosen reference depths $z_a = \{5, 10, 20, 30\}$\,mm of increased suppression.

The time-shift adjustment of $s_-(\vr,t)$ equals a movement of the on-axis zero-crossing depth coordinate of the $\tau(z)+\tau_a$, and consequently a change of depth where the gain-factor model, given in Eq.\:\ref{eq:gain_factor_def} predicts $G=0$ and hence zero SURF beam amplitude. However, the realized time-shift adjusted SURF beams display a beam amplitude at $z_a$ which gets less suppressed with increasing $z_a$. An explanation to this is that the effect of the LF manipulation is not solely a pure time-shift but also involves some pulse-form distortion, \emph{e.g.} {due to} pulse compression or expansion. This distortion may be caused by an experienced LF pressure which varies along the time axis for a HF pulse having propagated to a given spatial coordinate. 

The shown adjustments using a general filter $\hh$ generate full transmit beams suppression at the $z_a$ depths on-axis. However, the corresponding SURF difference HF pulse gets significantly distorted within the imaging region when $20{\lesssim}z_a{\lesssim}30$\,mm, as exemplified for $z_a = 30$\,mm in Fig.~\ref{subfig:bad_focalfield}, where the on-axis pulses are significantly less concentrated in time within the imaging region than for $z_a{<}20$\,mm. This effect is due to the edge waves arriving on-axis in this region, hence enforcing $\hh_{z_a}$ to be complicated in order to absolutely equalize $s_+(\vr,t)$ and $s_-(\vr,t)$. Longer difference pulses in the imaging region cause decreased range resolution. %

The determination of $\hh_{z_a}$ is here based solely on the on-axis fields. Hence the adjusted transmit-beam suppression may be insufficient off-axis. Optionally the calculation of $\hh_{z_a}$ could be based not only on the on-axis, but also the off-axis characteristics of $s_+(\vr,t)$ and $s_-(\vr,t)$. However, the adjusted transmit beams realized here and partially shown in the results section prove to have little energy off the beam axis for adjustments with $z_a{<}35$\,mm. 

The increased off-axis energy with increasing $z_a$ is reflected in Fig.~\ref{fig:Qza} where $Q_{z_a}$ is shown to decrease from ${\sim}75$\,dB at $z_a=0$ to ${\sim}0$\,dB at $z_a=30$\,mm for $\hh_{z_a}$ adjustment. The corresponding time-shift adjusted beams give a $Q_{z_a}$ decreasing from ${\sim}30$\,dB to $0$\,dB within the same depth interval, while the non-adjusted SURF beam at any $z_a$ depth gives a $Q_{z_a}{\sim}10$\,dB below what is attained using the time-shift adjustment. There is no significant gain in $Q_{z_a}$ by utilizing $\hh_{z_a}$ adjustment instead of time-shift adjustment for $z_a{\gtrsim}20$\,mm.

For the adjustments realized on the example dataset, the beams for $z_a{\gtrsim}30$\,mm suffer from giving high beam energy within the near-field. Thus the reduced sensitivity to reverberation-generating scatterers at $z_a$ is then attained at the cost of high sensitivity to reverberation-generating scatterers within the near-field. Since an imaging situation with strong reverberation-generating scatterers at great depths is likely to also have strong reverberation-generating scatterers within the near-field, the method thus has limited benefit compared to non-adjusted SURF when $z_a{\gtrsim}30$\,mm for the example dataset.

Determination of $\hh_{z_a}$ in a real imaging situation might be facilitated if \emph{à priori} information is known about how $s_+(\vr,t)$ and $s_-(\vr,t)$ differ depending on $z_a$. Potential sources for gathering of such information are: {(}i) received directly back-scattered HF signals from the $z_a$ depth, {(}ii) pre-recorded hydrophone measurements of the transmit waves at $z_a$, and {(}iii) numerical transmit wave field simulations, as utilized in this work. Based on either of these information sources, a set of pre-defined $\hh_{z_a}$ could be generated from within the scanner operator may choose at will. %

Fig.~\ref{fig:gain_factors} shows that application of a time-shift adjustment of opposite sign than for movement of the maximum suppression depth $z_a$ into greater depths, the gain factor instead grows within the whole near-field and the whole imaging region. The signal magnitude within the focal region is thus increased at the cost of less reverberation suppression. This may be favorable for the general image quality in case of little reverberation noise, and may be utilized to improve the sensitivity and signal-to-noise ratio (SNR), or may alternatively be traded-off into increased penetration depth. This is valid for situations including the one presented here, where the accumulated time-delay within the imaging region is smaller than the gain factor maximizing delay $1/(2 f_\HH)$. %

\begin{figure}[!tb]
\centering
\includegraphics[width=\figwidthCb]{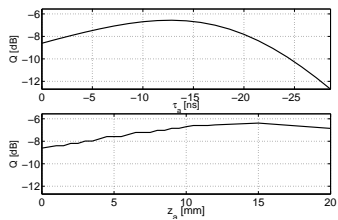}%
\caption{General beam quality ratio $Q$ as function of $\tau_a$ for the time-shift adjustment (top pane), and $z_a$ for the filter adjustment $\hh_{z_a}$ (bottom pane).}
	\label{fig:Q_total_z_a}
\end{figure}

As shown in Fig.~\ref{fig:gain_factors}, the relative change in gain-factor magnitude is larger within the near-field than within the focal region, for small time-shift adjustments $\tau_a$. The signal model predicts this effect to be amplified due to the lesser relative sensitivity to changes in the argument of the sinusoidal which models the gain factor, Eq.\:\ref{eq:gain_factor_def}, in case of $G$ being close to its maximum than when it is small. %
This observation inspires to assume that $\tau_a$ or $\hh_{z_a}$ adjustment in SURF beam generation might be useful not only to better suppress reverberations arising from a single depth, but also to get improved general reverberation suppression by decreasing the ratio between the total energy within the near-field $z\in[0,z_n]$ and the energy within the imaging region $z\in[z_n,z_f]$. This is illustrated in Fig.~\ref{fig:Q_total_z_a}, which shows the $\tau_a$ respectively $\hh_{z_a}$ influence on the general beam energy quality ratio $Q$, as defined in \cit{nasholm2009}: %
\begin{align}
	Q\triangleq {\displaystyle \sum_{z=z_n}^{z_f}\sum_{r=0}^{\infty}\sum_{\theta=0}^{2\pi} E(\myvecx)}\left/\ {\displaystyle \sum_{z=0}^{z_n} \sum_{r=0}^{\infty}\sum_{\theta=0}^{2\pi} E(\myvecx)}\right.,
\label{eq:quality_measure_general}
\end{align}
where $E(\myvecx)$ is the beam energy. %
Th{e attained} ratio $Q$, {as displayed in Fig.~}\ref{fig:Q_total_z_a}, has a maximum around the time-shift $\tau_a \approx 12.5$\;ns or the general filter $\hh_{z_a\approx 15\;\text{mm}}$, thus suggesting that such adjustments of $s_-(\vx,t)$ generate SURF beams generally more advantageous for reverberation suppression than without adjustment. The increase in general beam quality ratio is a modest ${\sim}1.5$\,dB. %

\section{Concluding remarks}
This work indicates that the SURF reverberation suppression transmit-beam may be adjusted to increased canceling of reverberation noise generated due to first scattering/reflection from a certain depth of choice $z_a$. The adjustment is done without change of the transmit SURF pulse complexes utilized, but instead by application of some filter, \emph{e.g.} a time-shift, on one of the propagated HF wave fields before SURF difference calculation. Different filters/time-shifts are utilized for different depths of first multiple scattering/reflection of the reverberation noise desired to suppress. %
The analysis of the Class\,I and II reverberation noise suppression abilities are based on synthetic SURF transmit-beams after adjustment. 

If the depth position $z_a$ of this first scattering/reflection to be suppressed is outside of the near-field, the reduction of sensitivity to scatterers at $z_a$ is attained at the cost of increased sensitivity to scatterers within the near-field. %

A strength of the method is that, as long as the receive HF imaging signals for propagation with both LF polarities are stored, there is no need for re-transmission of pulses to regulate the region of maximum reverberation suppression. 
The user may thus on-the-fly regulate the adjustment \emph{ad hoc}, while interactively observing which adjustment that generates the most clear image for the imaging region of interest depending on the nature of the reverberation noise present, or optionally to increase the sensitivity at large depths. The choice of adjustment may also be regulated to improve saved images as long as the HF receive IQ- or radio-frequency-data needed to form the SURF difference image is stored. The method thus gives a larger flexibility regarding regulation the reverberation suppression compared to \emph{e.g.} pulse inversion imaging where normally a new pulse couple has to be emitted if one aims to modify the transmit beam. %
%

The realization of the methods presented is limited to computer simulations of pulses in a homogeneous medium. Additional simulations for an inhomogeneous medium, as well as experimental setups with a real transducer and phantom, and \emph{in vivo} studies, are natural steps to further validate the feasibility in real imaging situations. %


\bibliographystyle{IEEEtran}


\begin{IEEEbiography}[%
	{\includegraphics[width=1in,height=1.25in,clip,keepaspectratio]{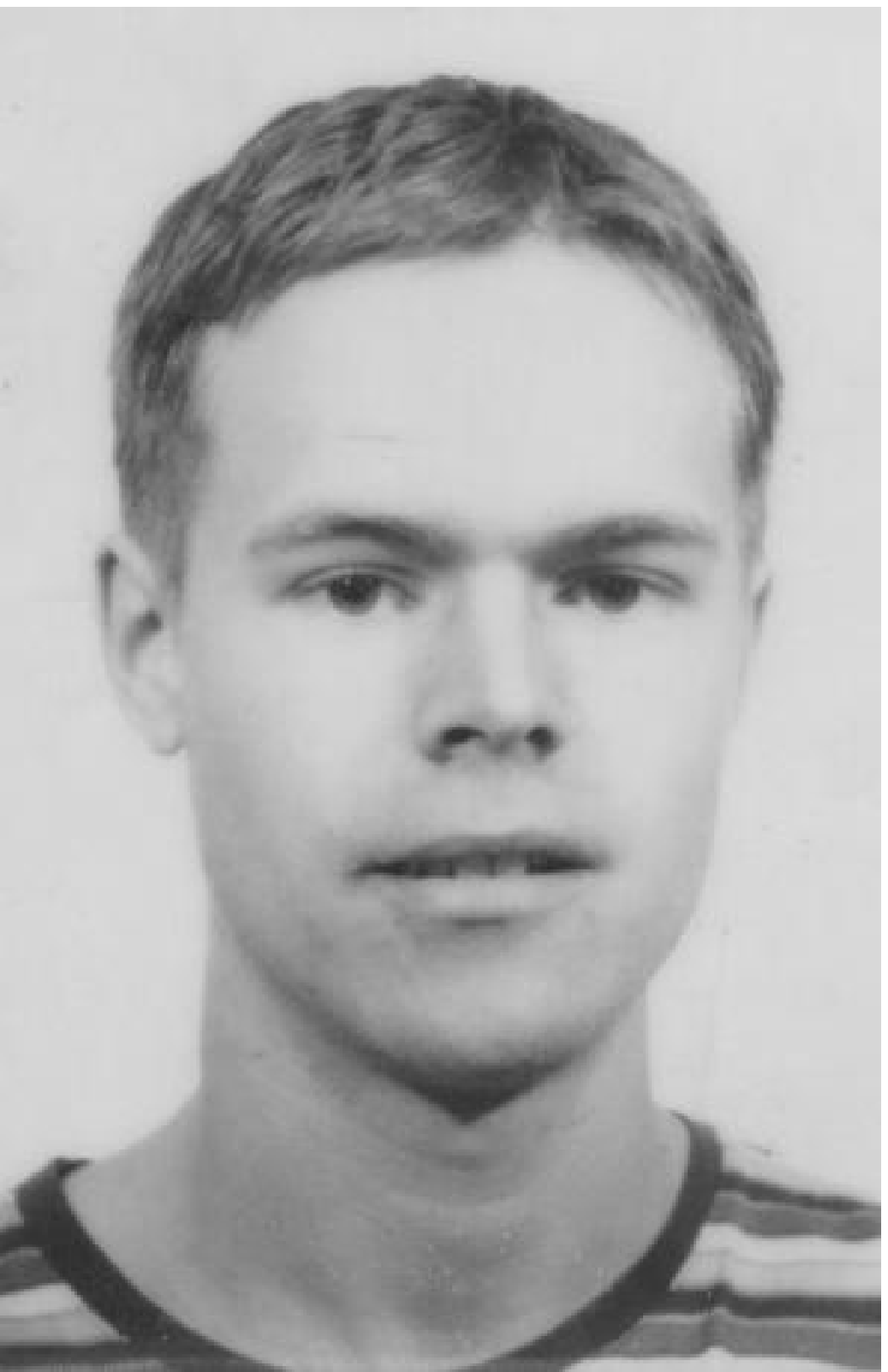}}]{Sven Peter Näsholm} 
was born in Örnsköldsvik, Sweden, in 1975. He received his M.Sc. degree in engineering physics in 2002 from Umeå University, Sweden. %
In 2008, he successfully defended his Ph.D. thesis entitled ``Ultrasound beams for enhanced image quality'' at the Norwegian University of Science and Technology, Trondheim, Norway. %
In 2009, he joined the Digital Signal Processing and Image Analysis group as a pos-doctoral fellow at the Department of Informatics, University of Oslo, Norway. %
His research interests are within the fields of sonar and ultrasound imaging including nonlinear effects, transducer design, field simulation, and acoustic noise suppression.

\end{IEEEbiography}
\begin{IEEEbiography}[%
	{\includegraphics[width=1in,height=1.25in,clip,keepaspectratio]{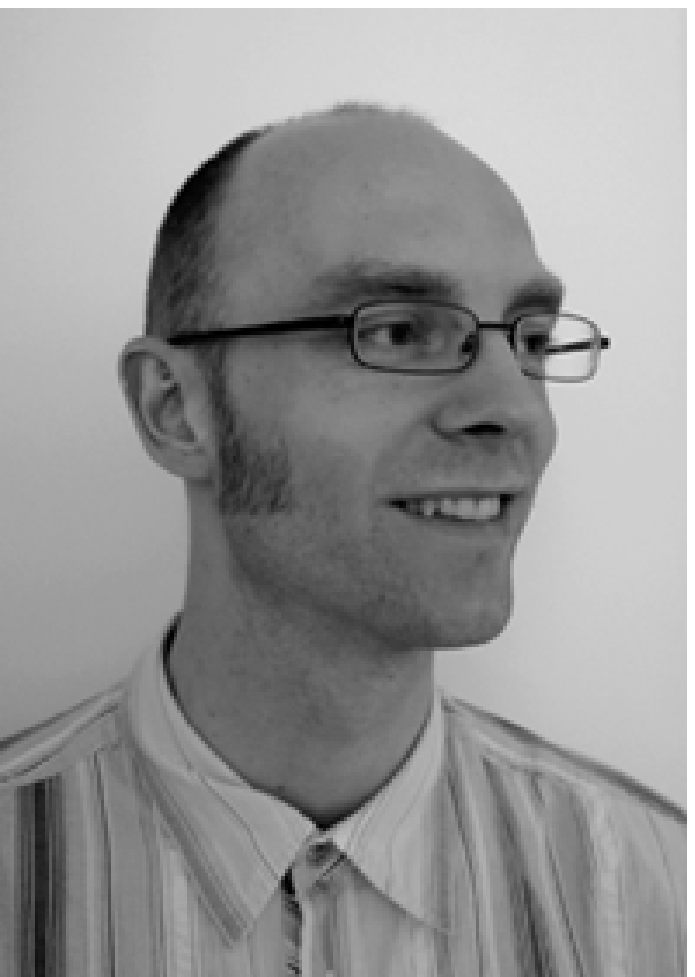}}%
]{Rune Hansen} 
was born in Horten, Norway in 1974. He earned his M.Sc. degree from the Department of Mechanics, Thermodynamics, and Fluid Mechanics at the Norwegian University of Science and Technology (NTNU) in 1998 and his Ph.D. degree from the Department of Engineering Cybernetics at the Faculty of Information Technology, Mathematics, and Electrical Engineering, NTNU, in 2004. In 1999 he joined the Department of Circulation and Medical Imaging at the Faculty of Medicine, NTNU, where he currently is employed in a part-time position as a Research Scientist.  Since 2006 he is also employed as a Research Scientist at the Department of Medical Technology, SINTEF Health Research, in Trondheim, Norway. His research interests include acoustics and signal processing with special focus at nonlinear acoustics and ultrasound contrast agents.    
\end{IEEEbiography}
\begin{IEEEbiography}[%
{\includegraphics[width=1in,height=1.25in,clip,keepaspectratio]{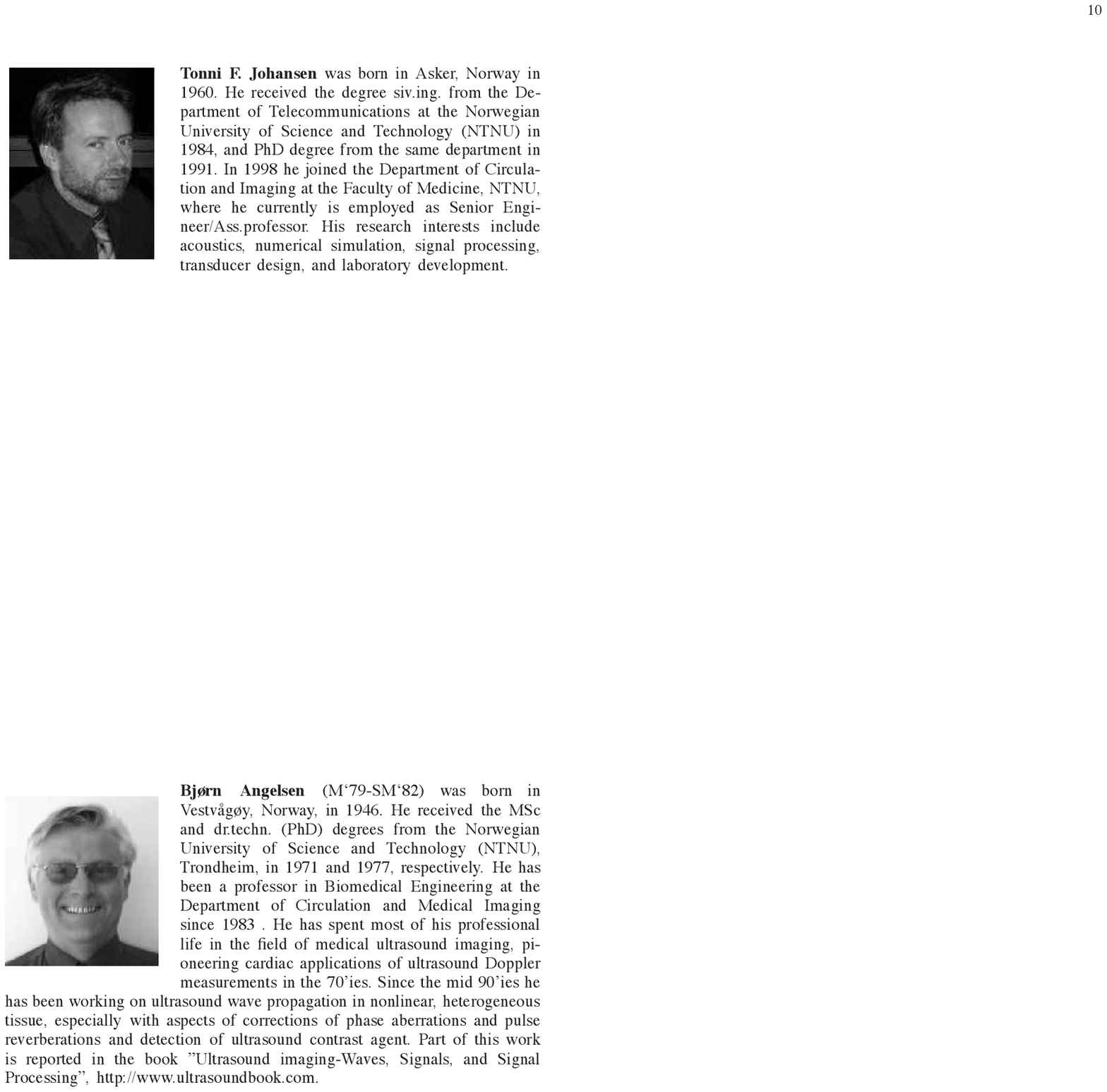}}]{Bjørn A. J. Angelsen}
was born 1946 in Vestvågøy, Norway. He received a MEE in 1971 from the Norwegian University of Science and Technology, Trondheim, and a Ph.D. from the same University in 1977. His Ph.D. work was on Doppler ultrasound measurement of blood velocities and
flow in the heart and the large arteries. He is a professor of Medical Imaging at the same university since 1983. In 1977--78 he was a visiting Post Doc at University of California, Berkeley, and Stanford Research Institute, Palo Alto. He has been strongly involved in development of cardiac ultrasound imaging instruments in collaboration with Vingmed Ultrasound, now GE Vingmed Ultrasound. He has written textbooks on ultrasound cardiac Doppler measurements and theoretical ultrasound acoustics, and holds several patents in the field of ultrasound imaging. 
\end{IEEEbiography}
\vfill

\end{document}